\documentclass[twocolumn]{aastex63}

\usepackage{amssymb}
\usepackage{amsmath}
\usepackage{epstopdf}
\usepackage{gensymb}
\usepackage{hyperref}
\usepackage{graphics,graphicx}
\usepackage{natbib}
\usepackage{multirow}
\usepackage{subfigure}

\def\lapp{\ifmmode\stackrel{<}{_{\sim}}\else$\stackrel{<}{_{\sim}}$\fi}
\def\gapp{\ifmmode\stackrel{>}{_{\sim}}\else$\stackrel{>}{_{\sim}}$\fi}

\def\frb200428{{FRB~20200428D}{\ignorespaces}}

\newcommand{\commentout}[1]{\ignorespaces}
\newcommand{\utkarsh}[1]{\textcolor{purple}{(Utkarsh: #1)}}

\definecolor{mycolor}{rgb}{0.2, 0.3, 1}

\def\chimefrb{\textsc{CHIME/FRB}{\ignorespaces}}
\def\sgr1935{{\textsc{SGR 1935+2154}{\ignorespaces}}}
\def\jax{\texttt{JAX}{\ignorespaces}}

\def\chime{\textsc{CHIME}{\ignorespaces}}
\def\bftheta{{\bm{\theta}}}
\def\bfdata{{\bm{D}}}

\mathchardef\mhyphen="2D

\begin{document}
\title{Comprehensive Bayesian analysis of FRB-like bursts from \sgr1935 observed by \chimefrb}
\shorttitle{\sgr1935 radio burst analysis}
\author[0000-0001-5553-9167]{Utkarsh Giri}
    \affiliation{Department of Physics, University of Wisconsin-Madison, 1150 University Ave, Madison, WI 53706, USA}
    \affiliation{Perimeter Institute for Theoretical Physics, 31 Caroline Street N, Waterloo, ON N25 2YL, Canada}
\author[0000-0001-5908-3152]{Bridget C. Andersen}
\affiliation{Department of Physics, McGill University, 3600 rue University, Montr\'eal, QC H3A 2T8, Canada}
\affiliation{Trottier Space Institute, McGill University, 3550 rue University, Montr\'eal, QC H3A 2A7, Canada}
\author[0000-0002-3426-7606]{Pragya Chawla}
\affiliation{Anton Pannekoek Institute for Astronomy, University of Amsterdam, Science Park 904, 1098 XH Amsterdam, The Netherlands}
\author[0000-0002-8376-1563]{Alice P. Curtin}
\affiliation{Department of Physics, McGill University, 3600 rue University, Montr\'eal, QC H3A 2T8, Canada}
\affiliation{Trottier Space Institute, McGill University, 3550 rue University, Montr\'eal, QC H3A 2A7, Canada}
\author[0000-0001-8384-5049]{Emmanuel Fonseca}
\affiliation{Center for Gravitational Waves and Cosmology, West Virginia University, Chestnut Ridge Research Building, Morgantown, WV 26505, USA}
\affiliation{Department of Physics and Astronomy, West Virginia University, P.O. Box 6315, Morgantown, WV 26506, USA}
\author[0000-0001-9345-0307]{Victoria M. Kaspi}
\affiliation{Department of Physics, McGill University, 3600 rue University, Montr\'eal, QC H3A 2T8, Canada}
\affiliation{Trottier Space Institute, McGill University, 3550 rue University, Montr\'eal, QC H3A 2A7, Canada}
\author[0000-0001-7453-4273]{Hsiu-Hsien Lin}
\affiliation{Institute of Astronomy and Astrophysics, Academia Sinica, Astronomy-Mathematics Building, No. 1, Sec. 4, Roosevelt Road, Taipei 10617, Taiwan}
\affiliation{Canadian Institute for Theoretical Astrophysics, 60 St.~George Street, Toronto, ON M5S 3H8, Canada}
\author[0000-0002-4279-6946]{K.~W. Masui}
\affiliation{MIT Kavli Institute for Astrophysics and Space Research, Massachusetts Institute of Technology, 77 Massachusetts Ave, Cambridge, MA 02139, USA}
\affiliation{Department of Physics, Massachusetts Institute of Technology, 77 Massachusetts Ave, Cambridge, MA 02139, USA}
\author[0000-0003-3154-3676]{Ketan R. Sand}
\affiliation{Department of Physics, McGill University, 3600 rue University, Montr\'eal, QC H3A 2T8, Canada}
\affiliation{Trottier Space Institute, McGill University, 3550 rue University, Montr\'eal, QC H3A 2A7, Canada}
\author[0000-0002-7374-7119]{Paul Scholz}
\affiliation{Dunlap Institute for Astronomy \& Astrophysics, University of Toronto, 50 St.~George Street, Toronto, ON M5S 3H4, Canada}
\affiliation{Department of Physics and Astronomy, York University, 4700 Keele Street, Toronto, Ontario, ON MJ3 1P3, Canada}

\author[0000-0001-5002-0868]{Thomas C. Abbott}
\affiliation{Department of Physics, McGill University, 3600 rue University, Montr\'eal, QC H3A 2T8, Canada}
\affiliation{Trottier Space Institute, McGill University, 3550 rue University, Montr\'eal, QC H3A 2A7, Canada}

\author[0000-0002-3615-3514]{Mohit Bhardwaj}
\affiliation{Department of Physics, Carnegie Mellon University, 5000 Forbes Avenue, Pittsburgh, 15213, PA, USA}
\author[0000-0003-4098-5222]{Fengqiu Adam Dong}
\affiliation{Department of Physics and Astronomy, University of British Columbia, 6224 Agricultural Road, Vancouver, BC V6T 1Z1 Canada}
\author[0000-0002-3382-9558]{B. M. Gaensler}
\affiliation{Dunlap Institute for Astronomy \& Astrophysics, University of Toronto, 50 St.~George Street, Toronto, ON M5S 3H4, Canada}
\affiliation{David A.~Dunlap Department of Astronomy \& Astrophysics, University of Toronto, 50 St.~George Street, Toronto, ON M5S 3H4, Canada}
\affiliation{Present address: Division of Physical and Biological Sciences, University of California Santa Cruz, Santa Cruz, CA 95064, USA}
\author[0000-0002-4209-7408]{Calvin Leung}
\affiliation{MIT Kavli Institute for Astrophysics and Space Research, Massachusetts Institute of Technology, 77 Massachusetts Ave, Cambridge, MA 02139, USA}
\affiliation{Department of Physics, Massachusetts Institute of Technology, 77 Massachusetts Ave, Cambridge, MA 02139, USA}
\affiliation{NHFP Einstein Fellow}
\author[0000-0002-2551-7554]{Daniele Michilli}
\affiliation{MIT Kavli Institute for Astrophysics and Space Research, Massachusetts Institute of Technology, 77 Massachusetts Ave, Cambridge, MA 02139, USA}
\affiliation{Department of Physics, Massachusetts Institute of Technology, 77 Massachusetts Ave, Cambridge, MA 02139, USA}

\author[0000-0002-3777-7791]{Moritz M\"unchmeyer}
\affiliation{Department of Physics, University of Wisconsin-Madison, 1150 University Ave, Madison, WI 53706, USA}
\author[0000-0002-8897-1973]{Ayush Pandhi}
\affiliation{David A.~Dunlap Department of Astronomy \& Astrophysics, University of Toronto, 50 St.~George Street, Toronto, ON M5S 3H4, Canada}
\affiliation{Dunlap Institute for Astronomy \& Astrophysics, University of Toronto, 50 St.~George Street, Toronto, ON M5S 3H4, Canada}
\author[0000-0002-8912-0732]{Aaron B. Pearlman}
\affiliation{Department of Physics, McGill University, 3600 rue University, Montr\'eal, QC H3A 2T8, Canada}
\affiliation{Trottier Space Institute, McGill University, 3550 rue University, Montr\'eal, QC H3A 2A7, Canada}
\author[0000-0002-4795-697X]{Ziggy Pleunis}
\affiliation{Dunlap Institute for Astronomy \& Astrophysics, University of Toronto, 50 St.~George Street, Toronto, ON M5S 3H4, Canada}
\author[0000-0001-7694-6650]{Masoud Rafiei-Ravandi}
\affiliation{Department of Physics, McGill University, 3600 rue University, Montr\'eal, QC H3A 2T8, Canada}
\affiliation{Trottier Space Institute, McGill University, 3550 rue University, Montr\'eal, QC H3A 2A7, Canada}
\author[0000-0001-6967-7253]{Alex Reda}
\affiliation{Department of Physics, Yale University, New Haven, CT 06520, USA}
\author[0000-0002-6823-2073]{Kaitlyn Shin}
\affiliation{MIT Kavli Institute for Astrophysics and Space Research, Massachusetts Institute of Technology, 77 Massachusetts Ave, Cambridge, MA 02139, USA}
\affiliation{Department of Physics, Massachusetts Institute of Technology, 77 Massachusetts Ave, Cambridge, MA 02139, USA}

\author[0000-0002-2088-3125]{Kendrick Smith}
\affiliation{Perimeter Institute for Theoretical Physics, 31 Caroline Street N, Waterloo, ON N25 2YL, Canada}
\author[0000-0001-9784-8670]{Ingrid Stairs}
\affiliation{Department of Physics and Astronomy, University of British Columbia, 6224 Agricultural Road, Vancouver, BC V6T 1Z1 Canada}
\author[0000-0002-9761-4353]{David C. Stenning}
\affiliation{Department of Statistics and Actuarial Science, Simon Fraser University, 8888 University Dr, Burnaby, BC V5A 1S6, Canada}
\author[0000-0003-2548-2926]{Shriharsh P.~Tendulkar}
\affiliation{Department of Astronomy and Astrophysics, Tata Institute of Fundamental Research, Mumbai, 400005, India}
\affiliation{National Centre for Radio Astrophysics, Post Bag 3, Ganeshkhind, Pune, 411007, India}
\affiliation{CIFAR Azrieli Global Scholars Program, MaRS Centre, West Tower, 661 University Ave, Suite 505, Toronto, ON, M5G 1M1 Canada}
  
\correspondingauthor{Utkarsh Giri}
\email{ugiri@wisc.edu}

\begin{abstract}
The bright millisecond-duration radio burst from the Galactic magnetar \sgr1935 in 2020 April was a landmark event, demonstrating that at least some fast radio burst (FRB) sources could be magnetars. The two-component burst was temporally coincident with peaks observed within a contemporaneous short X-ray burst envelope, marking the first instance where FRB-like bursts were observed to coincide with X-ray counterparts. In this study, we detail five new radio burst detections from \sgr1935, observed by the \chimefrb~instrument between October 2020 and December 2022. We develop a fast and efficient Bayesian inference pipeline that incorporates state-of-the-art Markov chain Monte Carlo techniques and use it to model the intensity data of these bursts under a flexible burst model. We revisit the 2020 April burst and corroborate that both the radio sub-components lead the corresponding peaks in their high-energy counterparts, respectively. For a burst observed in 2022 October, we find that our estimated radio pulse arrival time is contemporaneous with a short X-ray burst detected by \textsc{GECAM} and \textsc{HEBS}, and Konus-\textit{Wind} and is consistent with the arrival time of a radio burst detected by \textsc{GBT}. We present flux and fluence estimates for all five bursts, employing an improved estimator for bursts detected in the side-lobes. We also present upper limits on radio emission for X-ray emission sources which were within \chime's field-of-view at trigger time. Finally, we present our exposure and sensitivity analysis and estimate the Poisson rate for FRB-like events from \sgr1935 to be $0.005^{+0.082}_{-0.004}$ events/day above a fluence of $10~\mathrm{kJy~ms}$ during the interval from 28 August 2018 to 1 December 2022, although we note this was measured during a time of great X-ray activity from the source. This rate is informative for other future wide-field radio telescope detections of Galactic magnetars.
\end{abstract}


\section{Introduction}
\label{sec:intro}
Fast radio bursts are extremely bright transient radio pulses of very short ($\sim$ millisecond) duration, with dispersion measures (DMs) suggestive of extra-galactic origin. Discovered for the first time in 2007 \citep{Lorimer}, FRBs have emerged as one of the most exciting phenomena in the field of time-domain astronomy \citep{doi:10.1146/annurev-astro-091918-104501, Petroff:2019tty}. 
Explaining their origin has become a central unresolved problem in astronomy, and numerous theories have been proposed to explain the extraordinary emission mechanism that powers them and progenitors that host them \citep{Platts:2018hiy}. 

FRBs exhibit diverse phenomenology, adding richness and mystery to their possible origin mechanism. To date, thousands of FRBs have been detected \citep[e.g.][]{aab+21}
with a high fraction of them being one-off events where only a single component of emission from the source was observed. However, a small number of FRB sources have been seen to repeat, i.e.,\ the same source emits pulses intermittently \citep[e.g.][]{abb+19c,fab+20,abb+23}. A handful of these repeaters have also shown periodicity in their activity periods \citep{aab+20,abb+21b,css+21}. The repeaters and apparent non-repeaters also show statistically different morphologies \citep{Pleunis:2020vug, abb+19c, abb+23} with some repeaters having a puzzling downward-drifting structure in time-frequency phase space.

The sources of FRBs are still unknown. To understand them better, source localization and host association are key.  However, there is one case where we have detected FRB-like radio pulses from a previously known source:  the Galactic magnetar \sgr1935 \citep{abb+20,Bochenek:2020zxn}. While the sources of FRBs are still not definitively known, the detection of an FRB-like \footnote{{We call the radio bursts FRB-like as their spectral luminosity is several orders of magnitude greater than the pulsed emission seen from \sgr1935 \citep{Zhu:2023spq} and above that of a typical Giant pulse, but still a few orders of magnitude below that of FRBs.}} burst \frb200428 from \sgr1935 was the first direct evidence that magnetars could be a source of at least some FRBs.  Specifically, a radio burst with two sub-components was seen on 2020 April 28, with the sub-components separated by $\sim 30$ ms, following a period of unusually high X-ray activity. The radio sub-bursts were contemporaneous with hard X-ray emission peaks observed independently by {Konus-\textit{Wind}}, {INTEGRAL}, {Insight-HXMT} and {AGILE}~\citep{rsf+21, msf+20, Borghese:2020nxq, tcu+21}.

This landmark event and the associated high X-ray activity prompted several follow-up observations. On 2020 April 30, during the course of a month-long monitoring campaign, the Five-hundred-meter Aperture Spherical Telescope (FAST) observed a faint and highly polarized radio pulse from \sgr1935 which had a fluence of 51 mJy ms \citep{2020ATel13699....1Z}. Subsequently, \citet{Kirsten:2020yin} performed hundreds of hours of follow-up observations of \sgr1935  using several telescopes under a coordinated campaign, and detected two radio bursts on 2020 May 24 within a span of $\sim 1.4$s, with fluences of $112 \pm 22$ Jy ms and $24 \pm 5$ Jy ms, respectively. Unlike the April 2020 event however, none of these radio bursts was found to be accompanied with emission at X-ray wavelengths. %

On 2020 October 8, \chimefrb\  detected three more bursts from \sgr1935 within a span of $\sim$3 sec. The detection was rapidly communicated with the community \citep{g++20} and prompted a new follow-up observation campaign by the Five-hundred-meter Aperture Spherical Telescope (FAST). The campaign resulted in the detection of hundreds of periodic, sub-Jansky radio pulses, anti-aligned with the X-ray emission and likely of pulsar-mode origin, powered by spin-down energy \citep{Zhu:2023spq}. After a prolonged lull in activity of over 2 years, \chimefrb\ detected another FRB-like burst from \sgr1935 on 2022 October 14 \citep{d++22} and another one on 2022 December 1 \citep{p++22}. Interestingly, based on preliminary estimates of the arrival times, the October 14 burst was reported to be coincident with a short, 250 ms X-ray burst observed by GECAM and HEBS \citep{2022ATel15682....1W} and by Konus-\textit{Wind} \citep{2022ATel15686....1F} and also with a radio bursts detected by the Green Bank Telescope (\textsc{GBT}) at 5 GHz during a C-Band session \citep{2022ATel15697....1M}.

Studies of these events from \sgr1935 offer a unique opportunity to deepen our understanding of the transient universe. For example, comparing burst rates of regular FRBs with rates of FRB-like bursts from \sgr1935-like magnetars can tell us what fraction of FRBs might be coming from magnetars. Identifying coincident radio and X-ray bursts in space and time or putting upper limits on radio flux is an important step towards constraining FRB models. Moreover, given their special placement in the luminosity-duration energetics space, accurate flux measurements for \sgr1935 events and FRBs at large can offer critical insights.

The need for accurate and precise determination of burst parameters goes further than this. For example, a precise measurement of relative arrival time between a radio burst and an associated high-energy counterparts could help distinguish between different models of FRB emission \citep[e.g.][]{Margalit:2020luq,lp20}. In the case of \sgr1935, we already have two events where such an accurate time of arrival (ToA) estimation is needed. Estimating radio ToA correctly requires an adequate burst model together with a robust inference framework to disentangle propagation effects. More generally, as the FRB field develops, \commentout{we will have massive amounts of high resolution data revealing hard-to-model, complex burst morphologies. The} the FRB data quality as well as the count is expected to grow very rapidly\commentout{in the next few years}, enabling interesting science analysis via observables like pulse dispersion and scatter broadening. This necessitates the development of a fast, accurate and modular analysis framework that will enable processing large sets of data correctly while allowing for seamless integration of custom burst models for complex morphology. %

In this paper, we present an analysis of more  \sgr1935 radio bursts seen by \chimefrb~ using a fast and flexible Bayesian inference pipeline. After briefly describing the \chimefrb~ backend in \S\ref{sec:instrument} and our observations in \S\ref{sec:observations}, we describe the burst modeling in \S\ref{sec:pulse_model} . In \S\ref{sec:bayesian}, we describe our Bayesian inference pipeline and validate it using simulations. We analyze the intensity data of \sgr1935 events and present results in \S\ref{sec:results}. We reanalyze \frb200428 with a more flexible model and present an updated estimate of its arrival time. We confirm that both the radio sub-bursts of \frb200428 lead their contemporaneous high-energy counterparts and rule out their simultaneity at a high significance. In \S\ref{sec:flux_fluence}, we present flux and fluence estimates for all observed bursts. We employ an improved estimator for new bursts detected in the side-lobes. In \S\ref{subsec:radio_upper_limit}, based on non-detection of radio activity, we present upper limits on radio emission for X-ray pulses that were observed while their sources were within \chime's field-of-view. Finally in \S\ref{subsec:exposure_sensitivit}, we present our exposure and sensitivity analysis which is used to obtain the  rate of FRB-like bursts from \sgr1935.  Throughout the paper, our unit for DM are $\mathrm{pc~cm^{-3}}$ and time-like parameters are in milliseconds unless stated otherwise.

\section{Instrument}
\label{sec:instrument}

The Canadian Hydrogen Intensity Mapping Experiment (\textsc{CHIME}) is an interferometric telescope at the Dominion Radio Astrophysical Observatory in British Columbia, Canada. Designed initially with the aim of mapping neutral hydrogen in the universe \citep{Newburgh_2014, Bandura_2014, abb+22}, the instrument was supplemented
with a dedicated backend for real-time detection of FRBs --- \chimefrb~ \citep{abb+18}.
Here we summarize the \chimefrb~system. 

The \chime\ telescope comprises four 20-m $\times$ 100-m cylindrical reflectors with no moving parts, aligned in the North-South direction and operating between $400-800$ MHz. Each of these reflectors has 256 dual-polarization antennas installed along its focal plane axis. The analog signal received by the feeds gets digitized to a time-series with a 1.25~ns time-resolution before getting channelized to 1024 frequency channels at a resolution of  2.56 $\mu s$ using customized integrated circuits called Field-Programmable Gate Arrays (FPGAs). The channelized output from the FPGAs is processed by a series of pipelines. First among them is L0 pipeline which beamforms using Fast-Fourier Transforms \citep{Tegmark_2009,nvp+17,masui2019algorithms}. It is built out of 256 GPU nodes that correlate data from all the feeds with appropriate time delays to constructively interfere them to produce beams digitally ``pointed'' to a particular location in the sky. In total, L0 produces 1024 beams on the sky and further up-channelizes the data stream by a factor of 16 (yielding a total of 16384 radio-frequency channels) at a time resolution of 0.983 ms.

The data then go to the L1 pipeline, which is responsible for detecting FRBs. After filtering out terrestrial radio frequency interference \citep{Rafiei-Ravandi:2022rwl}, the CPU-based L1 pipeline uses a near optimal tree-based incoherent dedispersion algorithm \texttt{bonsai} to search for FRBs in real-time. The pipeline utilizes a ring-buffer to store intensity data, and upon detecting a candidate event above a nominal signal-to-noise (S/N), it triggers a dump of a few seconds of data around the event for further offline analysis. Finally, the pipeline also has a machine-learning based routine for sifting through the candidate events to identify genuine FRBs. Subsequently, the L2/L3 pipeline performs another iteration of RFI removal and groups together instances of an event that are detected in multiple beams. L2/L3 is also responsible for 
possible source identification and association. The last in the series is the L4 pipeline, which is responsible for maintaining a database for candidate meta-data and carrying out user-defined specific actions dependent on source identity.

\section{Observations}
\label{sec:observations}
\chimefrb~detected an extremely bright radio burst from the direction of \sgr1935 on 2020 April 28 with two sub-components at approximate UTC (topocentric, 400MHz) 14:34:33 \citep{2020ATel13681....1S, abb+20}. The burst was detected in the far side-lobe of the telescope at an hour-angle (HA)  of $\sim22$\degree. As is typical of extremely bright far side-lobe events, it was detected in many of the \chimefrb~beams.\footnote{For \chimefrb, events are in far side-lobes if they are
located at least several beam widths (1.3$^{\circ}$$-$2.6$^{\circ}$) away from the meridian.} The signature of such a detection is a dynamic spectrum (i.e., waterfall) with multiple regular ``spikes'' in radio frequency, as was seen for these events.
The second of the two sub-components was also detected by STARE-2 \citep{Bochenek:2020zxn} at 1281-1486 MHz. These radio observations occurred during a period of intense X-ray activity, in which hundreds of bursts were detected in the X-ray and $\gamma$-ray range. Several satellites reported X-ray bursts contemporaneous with the radio pulse \citep[e.g.,][]{msf+20,rsf+21,tcu+21}. A first analysis of the CHIME/FRB-detected radio bursts, including flux and fluence estimates, was presented in \cite{abb+20}.

On 2020 October 8, \chimefrb~detected three more bursts from \sgr1935 within a span of $\sim$3 seconds. All three bursts were detected in the main lobe of CHIME and the detection was reported via ATel including preliminary estimates of key burst parameters \citep{g++20}. After a gap of over 2 years, \chimefrb~ detected another burst from this source on 2022 October 14 at UTC 19:21:39.47 \citep{d++22} and another one on 2022 December 1 at UTC 22:06:59.08 at an HA of $-$99.8\degree~and $-$11.1\degree~from the \chime~ meridian, respectively \citep{p++22}. These bursts were detected in four beams and appeared to have spiky spectra indicative of far side-lobe detections. For all these events, our automated trigger system saved a few seconds of intensity data in the events vicinity for offline post-processing and analysis. For the bursts observed on 2020 October 14 and 2022 December 1 event, we were successful in saving a block of raw voltage data as well, an analysis of which will be presented elsewhere. In the next section, we describe the details of the model that we use to characterize the physical properties of observed bursts in the intensity data.   

\section{Burst Modeling}
\label{sec:pulse_model}
In this section, we describe our model for the radio burst intensity $I(\nu, t)$ that we fit to the dynamic spectrum $\mathcal{D}(\nu,t)$ of observed radio bursts. The basic pulse model we fit to an observed FRB accounts for its intrinsic features as well as propagation effects imprinted on the burst \citep{mls+15a, aab+21,McKinnon_2014}. The intrinsic profile of an FRB is modeled as a Gaussian pulse in time with a frequency-dependent modulation given by its spectrum. The Gaussian profile is a two parameter model given by
\begin{equation}
    I_{Gaussian} = \frac{1}{\sqrt{2\pi w^2}}\exp\Bigg(-\frac{(t-t_{0})^2}{2w^2}\Bigg), \nonumber
\end{equation}
where $t_0$ is the time of arrival of the burst at our telescope at a reference frequency (which we set to $\nu_{ref}=400.195$ MHz internally) and $w$ is the width of the pulse. We model the spectrum of the burst using a three-parameter power-law given by
\begin{equation}
    I_{spectrum} = A_s\Bigg(\frac{\nu}{\nu_0}\Bigg)^{n_s + \alpha \ln(\frac{\nu}{\nu_0})}, \nonumber
\end{equation}
where $A_s$ is the amplitude, $n_s$ is the spectral index and $\alpha$ models the running of spectral index as a function of frequency. They combine together to give us the total intrinsic pulse profile

\begin{equation}
     I_{intrinsic}(\nu, t) = I_{spectrum} * I_{Gaussian}. \label{eq:intrinsic}
\end{equation}

\noindent The resulting profile captures features that are likely intrinsic to the FRB generation process.  

In addition to these intrinsic features, the observed pulse is modified due to propagation effects. Chief among these are scattering and dispersion.
Scattering, or multi-path propagation of the FRB, causes many copies of the pulse to be superimposed with different delays. This results in a temporal broadening of the pulse which can be modeled under a thin screen approximation where the Gaussian profile gets convolved with a decaying exponential of the form $e^{-(t-t_0)/\tau_s}$, where the scattering time is given by
\begin{equation}
    \tau_s = \tau \bigg( \frac{\nu}{\nu_s}\bigg)^{SI},
    \label{eq:scattering}
\end{equation}
where $\tau$ is called the scattering 
time, is defined at a reference frequency $\nu_s$, and parameterizes the spectral broadening. We choose the center of the observing bandwidth as the reference frequency for scattering i.e. $\nu_s=600$ MHz. The parameter $SI$ is called the scattering index and characterizes the frequency dependence of the broadening. The scattering index depends on the nature of plasma inhomogeneities and in particular their power spectrum ($\propto \langle n_e^2 \rangle$). It can be a free parameter in principle but in the literature, its value is fixed to the theoretically motivated value of $-$4 or $-$4.4 \citep[e.g.][]{1971ApJ...164..249L,1977ARA&A..15..479R}. Thus,
\begin{equation}
{I_{scattered}(\nu, t) = I_{intrinsic}\otimes \Bigg[\frac{\theta(t-t_0)}{\tau_s} e^{-(t-t_0)/\tau_s} \Bigg ]} 
\end{equation}
where $\theta(t)$ is the Heaviside function and $\otimes$ denotes a convolution operation. 

Finally, the cold plasma dispersion leads to a frequency dependent delay in the arrival of the FRB pulse with a time shift given by
\begin{align}
    \Delta t(\nu) &= \frac{1}{K}\frac{DM}{\nu^{DI}} \label{dispersiondelay}
\end{align}
where the dispersion constant $K=241$ GHz$^{-2}$ cm$^{-3}$ pc s$^{-1}$ \citep{manchester_and_taylor} and $DM=\int n_e dl$. The integral over the electron density $n_e$ counts the number of electrons encountered by the FRB along its path from source to observer and is called the dispersion measure ($DM$). The parameter $DI$ is the dispersion index and captures the frequency dependence of the dispersion relation. The $DI$ parameter is typically set to $-$2 in most pulsar and FRB analyses. The leading order correction from this quadratic behavior is proportional to $({\nu_{plasma}}/{\nu_{emission}})^4$ and is highly suppressed. Incorporating this time shift, we have
\begin{equation}
    I(\nu, t)_{overhead} =   I_{scattered}(\nu, t - \Delta t({\nu})),
\end{equation}
where $\Delta t({\nu})$ is given by Eq.~(\ref{dispersiondelay}) and is dependent on two new parameters: $DM$ and $DI$.

Lastly, the intensity profile $I(\nu,t)_{overhead}$ gets attenuated by the beam response \footnote{\url{https://chime-frb-open-data.github.io/beam-model}} of the telescope as the signal is captured at the feed and processed. While fitting for the FRB parameters, we account for the attenuation by using models of the primary and the FFT-based synthesized beams. The primary beam is basically the beam response of a single feed over its cylinder. However, additional complex features appear on top of it due to instrumental effects like reflections and cross-talk between feeds. The primary beam for \chimefrb~is calibrated using data-driven holography and is accurate to within 10\%. A more detailed discussion of the primary beam can be found in \citet{abb+22}. The FFT-based synthesized beams, although complex and full of spatial and spectral features, are fully deterministic \citep{nvp+17}. Their effect can be accounted for up to the resolution of the beam. Incorporating the composite beam attenuation, the intensity data can be modeled as
\begin{equation}
    I(\nu, t) =   I_{overhead}*\mathcal{B}(Dec, RA) \label{eq:full_model}
\end{equation}
where $\mathcal{B}(Dec, RA)$ is the composite beam model at the known declination and right ascension of the \sgr1935. A discretized version of Eq.~(\ref{eq:full_model}) gives the final model we use for modeling the 2-d data array $\mathcal{D}(\nu_i,t_j)$ given arbitrary values of the parameters $\bftheta=(t_{0}, DM, w, A_s, n_s, \alpha, \tau, SI, DI)$. In our fitting, the position ($RA, Dec$) is kept fixed to the known sky position of the \sgr1935 given in degrees by $RA$ $=293.732\degree$ and  $Dec$ $=+21.897\degree$ \citep{ier+16}. %

For bursts with multiple components such as \frb200428, we model the intrinsic profile of each component independently while the propagation effects and instrument effects are modeled jointly. In particular, we expand the above prescription to model intrinsic features like width and spectrum independently for each of the additional components, whereas propagation effects like scattering and dispersion are fit using a common set of parameters for all the sub-components.

\section{Bayesian modeling}
\label{sec:bayesian}
Given the \emph{forward model} of an FRB as in Eq.~(\ref{eq:full_model}), the process of burst parameter estimation is the inverse problem of inferring the region in parameter space that agrees with the observed FRB data.

In order to infer the model parameters, we work under the probabilistic Bayesian framework which formalizes the inference problem. The Bayesian approach returns a distribution over the parameter space assigning a higher probability density to regions in the space which best fit the data. The approach naturally incorporates our previous beliefs or knowledge about the parameters expressed in the form of a distribution called the \emph{prior}. One of the advantages of the Bayesian formulation is the ability to consistently model and marginalize over parameters that one is not interested in but are still needed to describe the data. 

Formally, the posterior distribution $\mathcal{P(\bftheta|\bfdata)}$ over parameters $\bftheta=(t_{0}, DM, w, A_s, n_s, \alpha, \tau, SI, DI)$ of our fiducial model described in \S\ref{sec:pulse_model}, for a dataset $\mathcal{\bfdata}$ is given by
\begin{equation}
    \mathcal{P({\bftheta}| {\bfdata})} \propto \mathcal{P(\bfdata|\bftheta)P(\bftheta)} = \mathcal{L_{\bftheta}(\bfdata)P(\bftheta)} \\
\end{equation}
where $\mathcal{P(\bftheta)}$ is the joint prior distribution over the model parameters and $\mathcal{L_{\bftheta}(\bfdata)}$ is the likelihood. We assume prior independence and define $\mathcal{P(\bftheta)}$ as a product of independent distributions; either uniform or Gaussian distributions. Our likelihood function is a Gaussian noise model given by

\begin{equation}
    \mathcal{L_{\theta}} = \exp({-\chi^2(\bftheta, \mathcal{\bfdata})}/2) \label{eq:likelihood} \\
\end{equation}
with
\begin{align}
\ln \mathcal{L_{\theta}} = -\chi^2(\bftheta, \mathcal{\bfdata})/2= - \sum_{t,\nu} \frac{(\mathcal{\bfdata}(t,\nu) - I(t,\nu|\theta))^2}{2\sigma_{\nu}^2}
\end{align}
where $\sigma_{\nu}^2$ is the per channel variance of our data in the neighbourhood of the event.  

The posterior distribution over the parameters can be explored by obtaining an approximate sample using Monte Carlo Markov Chain (MCMC).

\subsection{Implementation}
\label{sec:implementation}
\subsubsection{Likelihood}
The likelihood defined in Eq.~(\ref{eq:likelihood}) is a numerically intensive function. For data $\mathcal{D}(\nu,t)$ of size ${16k \mbox{-}\mathrm{by}\mbox{-}\mathcal{O}(100)}$, the model ${I(t, \nu)}$, consists of  \commentout{$\mathrm{16k\mbox{-}by\mbox{-}m\mbox{-}by\mbox{-}\mathcal{O}(100)\mbox{-}by\mbox{-}n}$} ${16k \times m\times\mathcal{O}(100)\times n}$ elements, where $m$ and $n$ are factors by which model composition is upsampled in frequency and time with respect to the native data resolution. This property of the data lends itself to high throughput, single instruction multiple data (SIMD) computation. To leverage this property, we implement the model using the numerical library \texttt{JAX} \citep{jax2018github}. \texttt{JAX} is a modern high-performance library for numerical calculations on GPUs, implemented primarily for machine learning research and application. \texttt{JAX} uses XLA (Accelerated Linear Algebra; a domain-specific compiler for linear algebra) to compile and run programs on GPUs and has an API very similar to the popular numerical library NumPy. Compilation happens under the hood by default, with library calls getting just-in-time compiled and executed. But \texttt{JAX} also lets us compile our own Python functions into XLA-optimized kernels. We make use of this just-in-time compilation feature as well as efficient vectorization offered in \texttt{JAX} to optimize our likelihood code. This likelihood model has been used previously in several of our studies \citep[e.g.][]{abb+19a, abb+20}, but is re-written from scratch in \jax~ and integrated into the object-oriented approach of the pipeline.

\subsubsection{Sampling}
In order to further optimize our analysis, we implement a vectorized version of the popular affine-invariant MCMC sampling algorithm from scratch in \texttt{JAX}. The parallelized version of this  algorithm, first presented in \cite{foreman2013} along with a pure Python implementation, utilized disjoint sets of random walkers to sample the posterior distribution, where the walkers in one set had their position updated based on the walkers in the other set, making it amenable to SIMD principle. We implement the same basic idea in \texttt{JAX} and make it publicly available for general use\footnote{\url{https://github.com/utkarshgiri/jaims}}. 

Among the various known sampling approaches, and beyond the basic random-walk based samplers, there exists a class of samplers which utilize local gradient information of the distribution to efficiently converge to the target distribution. With \texttt{JAX}, one can precisely compute the gradient of a Python function relative to its parameters via the Automatic Differentiation algorithm. We leverage this feature to integrate the gradient-based No U-Turn Sampler (NUTS) \citep{neal2011mcmc, hoffman2011nouturn} into our repertoire as well.

For each of our burst analysis, described further in the next sub-section,  we use both the gradient-based NUTS sampler from the \texttt{numpyro} library and the native \texttt{JAX}-based version of the affine-invariant sampler. We find that both samplers give consistent results.

\subsubsection{MCMC pipeline} %

For a single component event, our \emph{default} analysis fits progressively complex burst models to an approximately dedispersed, $16k\mbox{-}\mathrm{by}\mbox{-}\mathcal{O}(100)$ size dynamic spectrum data around the burst in an iterative fashion. The analysis begins by MCMC fitting a Gaussian burst model with parameters $\bftheta=(t_{0}, DM, w, A_s)$ to band-summed intensity time-series. The initial estimates for $t_0$ and $DM$ come from the real-time detection pipeline. The width $w$ is initialized to 1 ms and $log_{10}(A_s)$ is initially set to a reasonable value of ${-4}$ in native data units. All these parameters get a broad uniform prior around their initial estimate.  We use a uniform prior of 1D radius ($ \mathrm{4, 8}$) in units of $\mathrm{ms}$ and $\mathrm{pc \; cm^{-3}}$ around the real-time detection estimates of $t_0$ and $DM$. This is a sufficiently generous prior given the accuracy of our real-time estimates. The uniform prior range for $w$ in ms is $\mathcal{U}(0, 10)$ and for $log_{10}(A_s)$ it is $\mathcal{U}(-10, 2)$.  We fix the $DI$ to $-2$. The posterior is then sampled using MCMC. This initial iteration refines the location of the burst in time-DM space and provides a good first estimate of the intrinsic pulse width along with a rough estimate of the amplitude. Using mean based point-estimates for $(t_{0}, DM, w, A_s)$ from this first MCMC iteration, we initialize our second iteration, where we additionally fit for the spectral parameters $n_s$ and $\alpha_s$, which have their initial values set to 0.  

\textbf{Basic Model:} In our third iteration of the MCMC fitting, we include scatter broadening by convolving the Gaussian pulse profile with a one-sided exponential as given by Eq.~(\ref{eq:full_model}). The additional parameters needed to model scattering are $\tau$ and $SI$ from Eq.~(\ref{eq:scattering}), but we fix $SI$ to $-4$ and fit only for $\tau$. Thus our model now has seven parameters in total; $\bftheta=(t_{0}, DM, w, A_s, n_s, \alpha, \tau)$ with $DI$ still kept fixed to $-2$ as before. Our initial estimates for $t_{0}, DM, w, A_s, n_s, \alpha$ come from the mean of the converged samples from the last iteration  while  $\tau$ is initialized to a reasonable value of 1 ms. We define a broad uniform prior around this initial value with a 1D radius of  (5, 2, 4, 4, 500, 500, 4) where time-like values are in ms. For e.g, our prior on time $t_0$ is $\mathcal{U}(\bar{t}_0 - 5, \bar{t}_0 + 5)$ where $\bar{t}_0$ is the mean estimate of $t_0$ from the last iteration of the fit. We call this model the `basic model' henceforth.

\textbf{Fiducial Model:} For the analysis presented in this work, our `fiducial model' is the full nine-parameter model given by Eq.~(\ref{eq:full_model}) which we fit for in the fourth and final iteration of the fitting process. Compared to the seven-parameter basic model, we now additionally fit for the parameters $DI$ and $SI$. The parameters $t_{0}, DM, w, A_s, n_s, \alpha$ and  $\tau$ are initialized to the mean value of MCMC samples obtained from the previous step. The two new free parameters $DI$ and $SI$ are initialized to their fiducial values of $-2$ and $-4$. We use flat priors for $t_{0}, DM, w, A_s, n_s, \alpha$ and  $\tau$ with 1d radius given by (5, 2, 3.16, 4, 200, 200, 1) where time-like values are in ms. We use a Gaussian prior with width of 0.001 for $DI$ and a width of 0.5 for $SI$. The burst model for this final iteration now also includes a treatment to model intra-channel dispersion smearing, as has been done in our previous works \citep{abb+19a}. We compose the model at an upsampled factor of 16 and 8 in time and frequency, respectively, which we then boxcar-convolve to get the predicted pulse at our instrumental resolution. The boxcar convolution serves to model the impulse response of our instrument. This additional step in the analysis increases our robustness to often ignored issues like dispersion smearing and instrument impulse modeling. This four-step iterative fitting procedure constitutes our default analysis.%

The burst models described above are the same as that used in \cite{mls+15a} and implemented in the \texttt{fitburst} burst fitting library \citep{fonsecafitburst}. An initial version of this MCMC pipeline was used for the Bayesian modeling of FRBs in \chimefrb's first science paper \citep{abb+19a}. The codebase has since been re-implemented from scratch in \texttt{JAX} with several optimizations incorporated along the way. The inference pipeline outlined above, with its relatively uninformative choice of prior has been successfully applied to hundreds of \chimefrb~events. However, some atypical events require special treatment. %
We describe these choices in detail in \S\ref{sec:results}.

\begin{figure*}[h!]
    \centering
    \includegraphics[width=0.9\textwidth, keepaspectratio,]{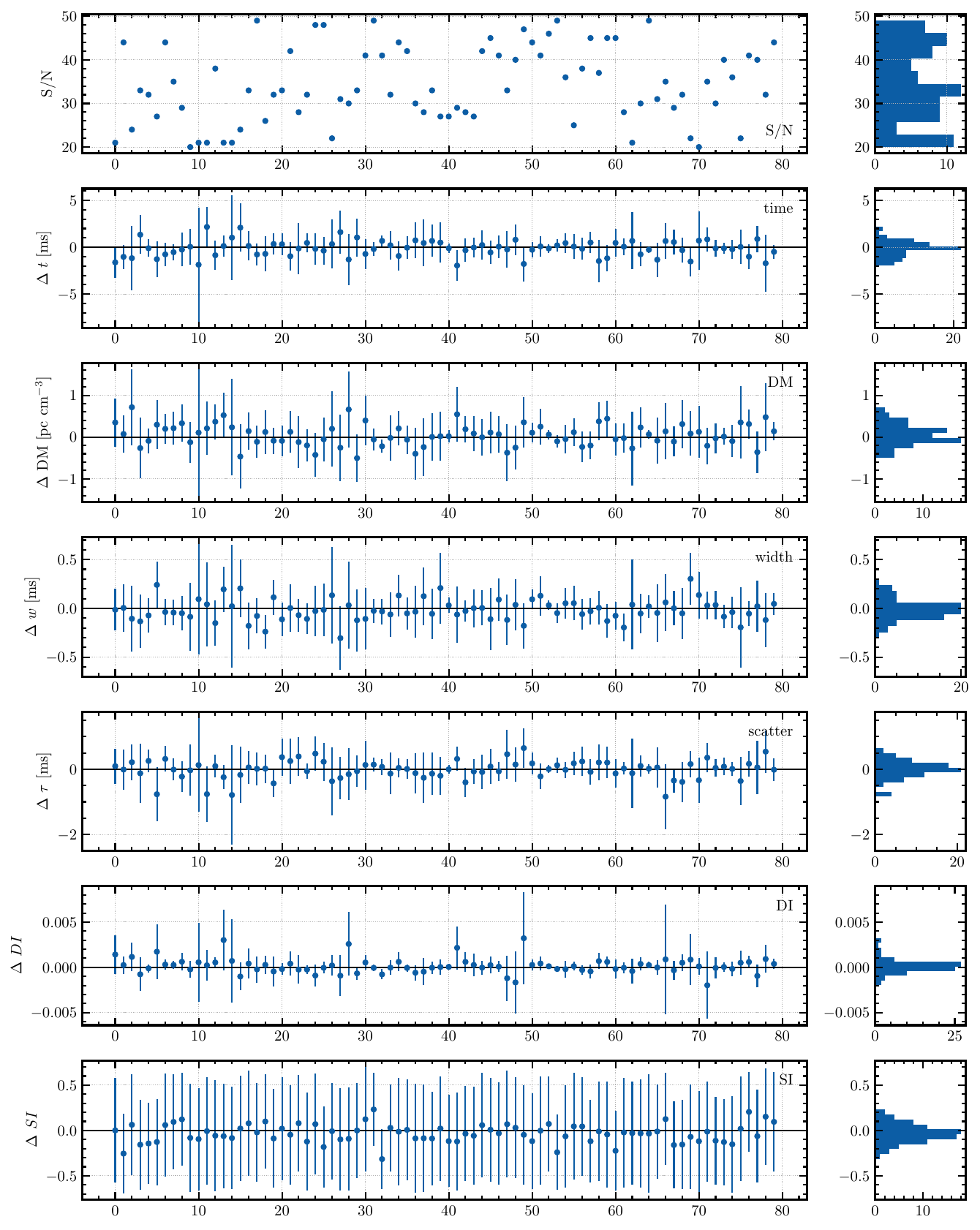}\caption{Results from validation study of our MCMC pipeline using simulated FRBs as described in \S\ref{sec:validation}. The study uses \texttt{simupulse} to simulate radio pulses with varying true physical parameters which are then fed to the MCMC pipeline for modeling and parameter estimation. The top panel shows S/N of the simulated burst. Remaining panels show difference between the true simulation values of parameters and their MCMC based median estimates for 80 independently simulated bursts along with the $2\sigma$ uncertainty. We use broad uniform priors on all the parameters and recover $\sim95\%$ of  the parameters within the uncertainty of $2\sigma$ and without any systematic bias. The study was performed on 200 simulated events but for the sake of clarity we show results for a subset of  simulated bursts. The spectral parameters $n_s$ and $\alpha_s$ were also found to be consistent with their true simulation values. }
    \label{fig:simulation}
\end{figure*}

\subsection{Validation}
\label{sec:validation}
Since the objective behind performing a comprehensive modeling of \sgr1935, and FRBs in general, is to obtain an accurate estimate of its parameters and their associated uncertainties, thoroughly testing our analysis pipeline is of paramount importance. We test the robustness of the MCMC pipeline by using it to model a suite of radio bursts simulated using high precision burst simulation library \texttt{simpulse}\footnote{\url{https://github.com/kmsmith137/simpulse/}} and with specifications of \chimefrb. For each simulation, we randomly and sparsely sample burst parameters from a hyper-space of $DM$, width and scattering. For simplicity, we keep remaining parameters of the bursts fixed to some fiducial value. In particular, the intrinsic spectrum is assumed to be broadband and flat ($n_s=0$, $\alpha_s=0$) and the $DI$ and $SI$ are fixed to $-2$ and $-4$, respectively. We simulate an FRB for the sampled parameter and add Gaussian noise and normalize it to a randomly chosen S/N in the range of 20 to 50. The simulated time-frequency data are seamlessly fed to the MCMC pipeline. The initial guess for time and $DM$ parameters needed by the MCMC pipeline are obtained by adding Gaussian perturbation to the true, known values of the parameters used for simulation. We follow this prescription to simulate and model 200 bursts. In our post-processing, we find that $\sim 95\%$ of each of the recovered parameters lie within $2\sigma$ of the true values used for simulation. None of the parameters show any kind of systematic bias. In Figure \ref{fig:simulation}, we show the fidelity of recovered parameters compared to true simulation parameters. We note that the validation analysis presented here is simple and idealized where the noise and RFI are not fully representative of what is encountered in reality.  

\begin{figure*}[h!]
    \centering
    \includegraphics[scale=0.5]{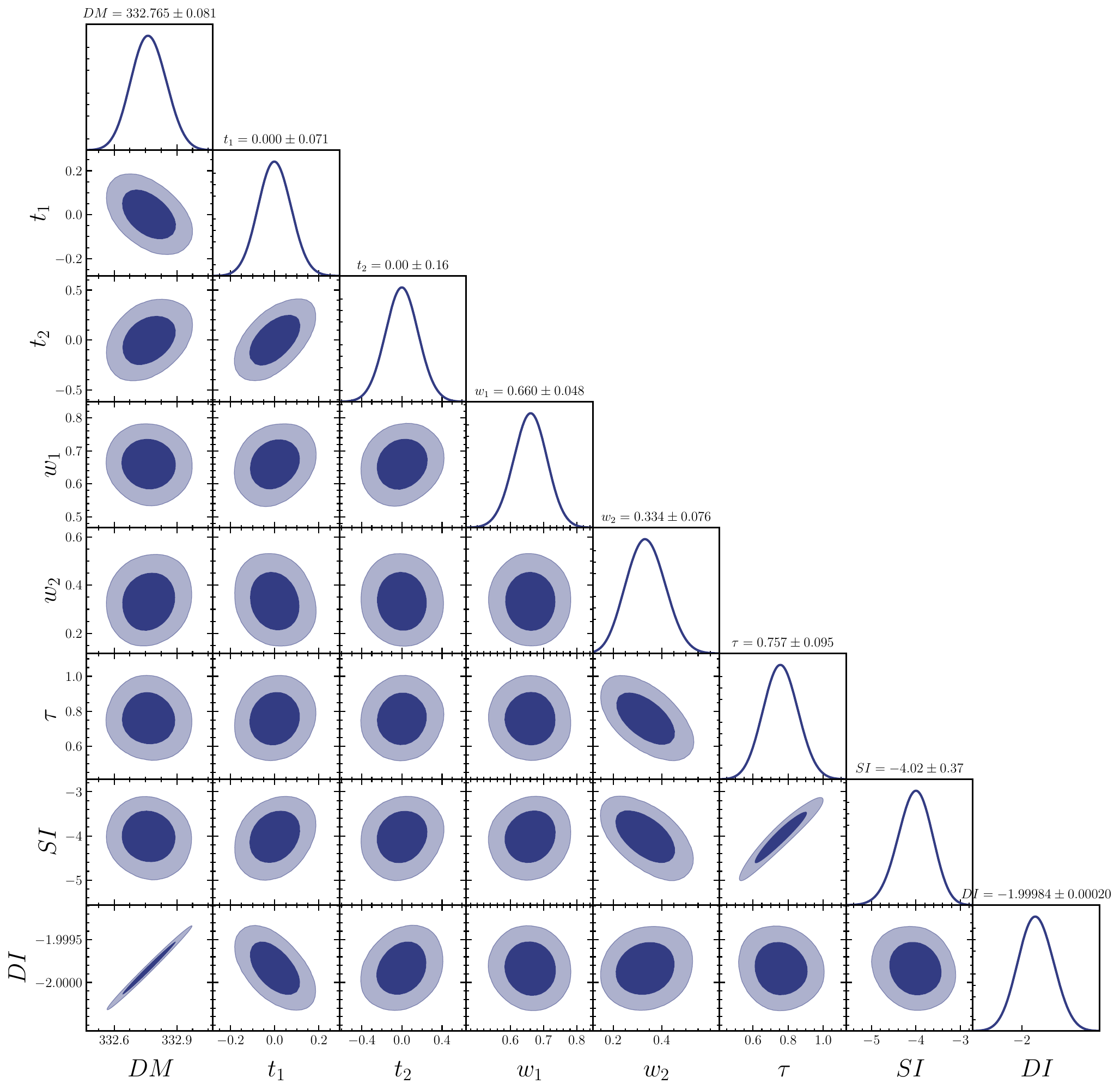}
    \caption{Triangle plot showing the two-dimensional joint posterior distributions of parameters as well as marginalized posterior distribution for a subset of parameters of our fiducial burst model for \frb200428. The modeling shown here is performed on data for \frb200428. The two sub-bursts are fit jointly using a common set of dispersion and scattering parameters as described in \S\ref{sec:implementation} and \S\ref{subsubsec:frb20200428}.  All time measurements are in milliseconds while  DM is in $\mathrm{pc~cm^{-3}}$. The values of $t_1$ and $t_2$ are standardized with respect to arrival time of the sub-bursts at 400.195 MHz. The plot is produced using \texttt{getdist} \citep{lewis2019getdist}. }
    \label{fig:triangle_plot}
\end{figure*}

\section{Results}
\label{sec:results}
\subsection{Burst modeling and Inference}
\subsubsection{\frb200428}
\label{subsubsec:frb20200428}
A first analysis of \frb200428 was published in \citet{abb+20}. The arrival times of the two sub-bursts were found to lead the contemporaneous/corresponding peaks in their high-energy counterpart light-curves by a few milliseconds. Establishing their simultaneity or their order of arrival is important for discriminating among magnetar and FRB emission models. A conclusive result would also guide future model building attempts. We therefore revisit the analysis using our newly developed MCMC pipeline. For the multi-component \frb200428, we model the two components of the bursts jointly, where the intrinsic parameters are fit independently while propagation effects are fit using a common set of parameters for both the sub-components. Thus we have a pair of estimate for the arrival time ($t_1$ and $t_2$) and pulse width ($w_1$ and $w_2$) corresponding to the two components but a single estimate for the propagation effects $DM$, $\tau$, $SI$ and $DI$.

For multi-beam detections, we have generally analyzed and presented results for the highest S/N beam data, as was the case in \citet{abb+20}. But for the extremely bright \frb200428, our burst model is inadequate in characterizing the highest S/N beam data and our Gaussian likelihood assumption breaks down. We do obtain reasonable results for each of the first 3 steps of the iterative fitting, including for the basic model (which is effectively the same as the fiducial model with sharply peaked priors on $DI$ and $SI$). However, once we fit the full nine-parameter fiducial model with relatively weak priors on model parameters as described in \S\ref{sec:implementation}, the inadequacies become apparent. The $DI$ retrieved after fitting the fiducial model to the data from the highest S/N beam is several sigmas away from the expected value of $-2$. Although such a discrepancy could point to interesting physics, we believe the deviation is possibly due to the failure of the Gaussian likelihood model for this high S/N, non-Gaussian event. The band-attenuated dynamic spectrum of \frb200428 shows a complex structure near the bottom of the band which is only made worse by the polyphase filter bank (PFB) leakage \citep{ple21}. Such a structure, which is not accurately described by our burst model, can lead to covariant shifts in the estimated values of the highly correlated parameter pairs $DM$ and $DI$.  %

To circumvent this issue, we make the conservative decision to analyze data and present results from a different beam that has a much lower detection S/N. To decide which beam to analyze out of the available options, we look for the beam with highest S/N detection for which the PFB leakage is not visually apparent in the dynamic spectra plot. We find that beam 2068 satisfies this criteria and so we re-run our fiducial analysis on its intensity data.

In Figure \ref{fig:triangle_plot}, we present two-dimensional joint posterior distributions of parameters for \frb200428 and report the best-fit parameter values in Table \ref{tab:properties}. The best-fit parameters are reasonably consistent with those reported in \citet{abb+20} particularly when we consider that the data, their modeling as well as the RFI masking strategy employed here are different compared to those of \citet{abb+20}. Crucially, the qualitative conclusions made about the arrival time of the sub-bursts remain unchanged. Both the radio sub-bursts temporally lead their corresponding high-energy counterparts. 

\begin{figure*}[ht!]
    \centering
    \includegraphics[width=0.96\textwidth]{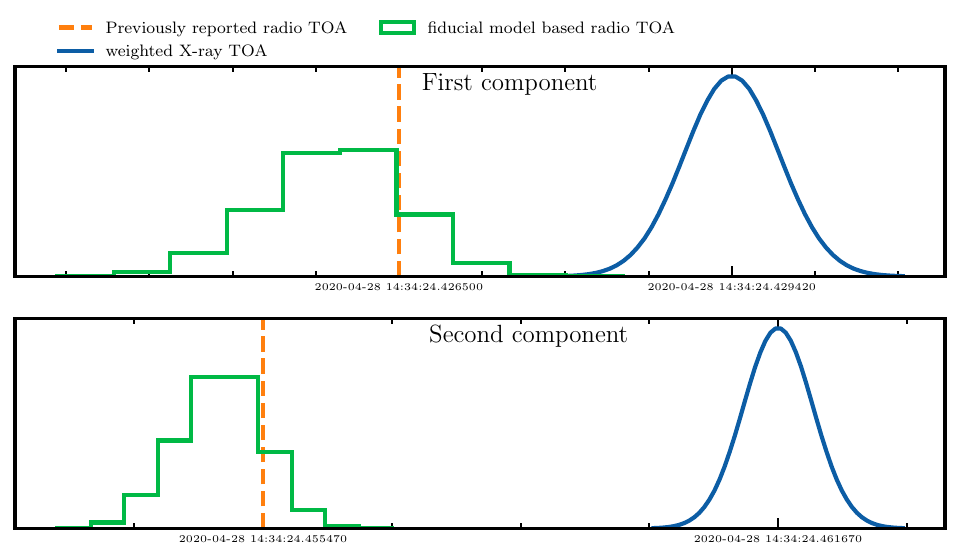}
    \caption{Estimated time of arrival (ToA) in UTC, geocentric for the two sub-components of \frb200428. The histogram in green is the marginalized arrival time for the radio pulse under our fiducial model which has scattering index and dispersion index, among others, as free parameters.  The orange dashed vertical line is the radio arrival time reported by \cite{abb+20}. The blue curve denotes the weighted arrival time of the near-coincident X-ray peak estimated by \cite{Ge:2023qke}. All the timestamps are at infinite frequency and include a correction for the dispersion delay for the radio pulse. }
    \label{fig:arrival}
\end{figure*}

\textbf{Quantifying the arrival time offset between radio sub-bursts and X-ray peaks: }  We use the approach and values reported in \citet{Ge:2023qke} to make quantitative statements about the offset between the arrival time of the two radio bursts and the corresponding peaks seen in the X-ray light-curve of several instruments. The \emph{weighted} X-ray arrival time of the first and second peaks, based on measurements made by {\it Insight-HXMT} High Energy, {\it Insight-HXMT} Medium Energy, { Konus-\textit{Wind}} and {INTEGRAL} in UTC, geocentric is 14:34:24.42942$\pm$0.00043 and UTC 14:34:24.46167$\pm$0.00043, respectively \citep{Ge:2023qke}. {The ToA estimate for the observed peaks in {\it Insight-HXMT} High Energy, {\it Insight-HXMT} Medium Energy and INTEGRAL come from a multi-Gaussian profile fit to the data in \citet{Ge:2023qke}, \citet{Li_2021}, and \citet{msf+20}. For the ToA from the Konus-$Wind$ light-curve data, the starting time of the 4-ms bin associated with the two \emph{wide} peaks are reported by \cite{Ridnaia:2020gcv}. To get a refined estimate of the ToA from the Konus-$Wind$ light-curve data, Gaussian profiles were fit to the publicly available data in \cite{Ge:2023qke}. \footnote{The result from this Gaussian model fit agrees reasonably well with a FRED model fit performed subsequently by the Konus-$Wind$ team (Ridnaia et al. (2023, private communication)).}  Based on these ToA estimates from Gaussian model fits,} the \emph{weighted} average delay of X-ray arrival time for the first peak with respect to the arrival time of first radio sub-burst reported in \citet{abb+20} is $\Tilde{\Delta}_1 = 2.92 \pm 0.43$ ms (refer to Table 3 of \citealt{Ge:2023qke}). The weighted uncertainty is dominated by the measurement made by {\it Insight-HXMT} High Energy which has the least uncertainty. Our refined estimate for the geocentric ToA at infinite frequency for the two radio components are UTC 14:34:24.42595$\pm$0.00061 and UTC 14:34:24.45497$\pm$0.00051, respectively. With our refined estimate, the delay is updated to $\Delta_1=3.47\pm0.73$ ms where the final uncertainty comes from adding radio and X-ray timing uncertainties in quadrature. Thus, {under these modelling choices}, we rule out simultaneity in the arrival times for the first peak of the radio burst and the contemporaneous high-energy counterparts at $4.7\sigma$. Similarly, for the second radio sub-burst and the corresponding high-energy counterpart, we find  $\Delta_2=6.70\pm0.67$ ms and thus rule out simultaneity of the two at $10\sigma$.  These offsets are many times the radio burst widths, many times the widths of the X-ray peaks but less than the overall X-ray burst envelope (see Fig. 1 in \citealt{msf+20}), and much less than the 3.24-s period of the magnetar. In Figure \ref{fig:arrival}, the arrival time estimates of radio and X-ray pulses are shown in UTC (geocentric).

\commentout{\textbf{Quantifying the offset between \emph{peak separation}:} Another very interesting quantity to look at is the temporal separation between the two peaks in radio and X-ray band. This quantity, being a \emph{differential} measurement, is free from many plausible systematic biases by design and hence would allow one to make robust claims about emission mechanisms. Based on the estimates reported above, the offset between peaks in the X-ray band is $32.25 \pm 0.61$ ms. The offset between radio sub-bursts is $29.02 \pm 0.80$ ms. Thus, at $3.2\sigma$ we can say that the peaks in the two energy bands have different time separation.}

We emphasize again that the radio ToA estimates are obtained using a very conservative choice of data and burst modeling where the choices were made \emph{a-priori} and were mainly dictated by the desire to robustly quantify the radio burst ToA. While we have not shown results for the highest S/N beam data due to its subpar fit quality, the offset in arrival time between the radio and X-ray bands is present even in that case, and is notably more significant than the conservative result we have presented.
{Finally, we note that our offset estimate and its significance analysis rely on the estimate of the X-ray arrival time published by several other groups, utilizing different datasets, modeling, and systematic mitigation strategies. The time resolution of the X-ray light-curve data and its binning, as well as the energy window of these datasets and dead-time correction strategies, are different for each group. Therefore, systematic biases to the individual timing estimates cannot be ruled out. More details about the data and modelling can be found in \cite{msf+20, Ge:2023qke, Li_2021, Ridnaia:2020gcv}. We have aggregated these estimates using inverse variance weighting and do not attempt to analyze them ourselves.} 

\begin{figure*}[h!]
    \centering
    \subfigure{\includegraphics[width=0.41\textwidth]{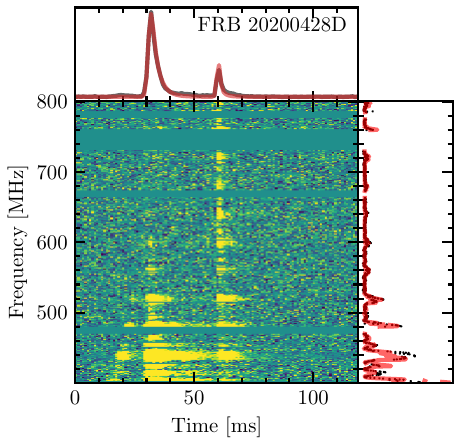}}
    \subfigure{\includegraphics[width=0.41\textwidth]{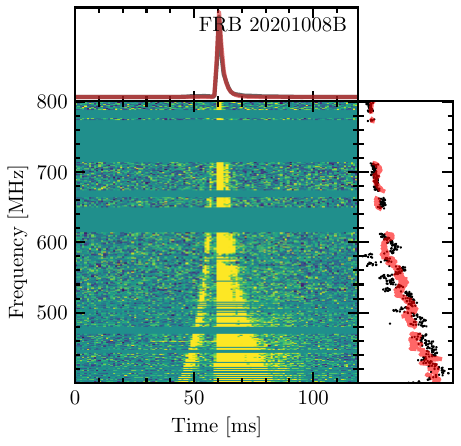}}
    \subfigure{\includegraphics[width=0.41\textwidth]{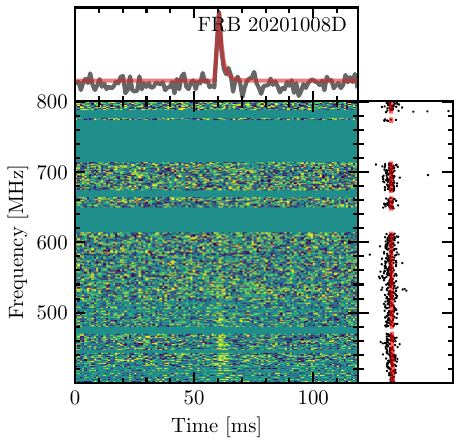}}
    \subfigure{\includegraphics[width=0.41\textwidth]{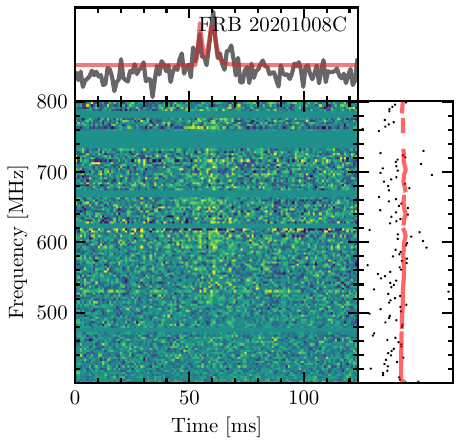}}
    \subfigure{\includegraphics[width=0.41\textwidth]{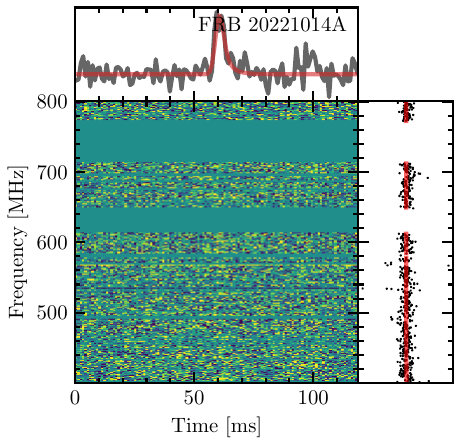}}
    \subfigure{\includegraphics[width=0.41\textwidth]{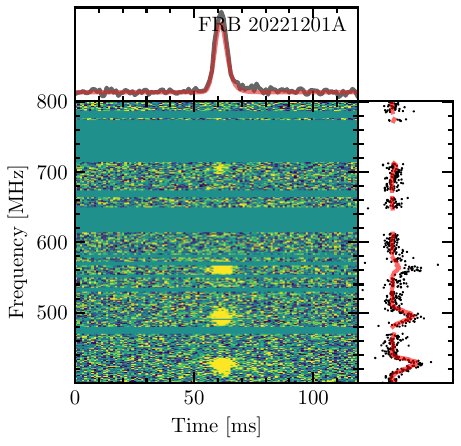}}

    \caption{Dynamic spectra of all the \sgr1935 radio bursts modeled in this paper. The top panel of each subplot shows de-dispersed and band-averaged data (grey) along with the model curve (red). The right panel of each sub-plot shows the time-averaged spectrum of the burst along with the beam-convolved model spectrum. The data correspond to the beam with the highest detection S/N and we use a customized mask for each spectrum. The curves around the brighter events are characteristic features resulting from spectral leakage in the polyphase filterbank \citep{ple21}}%
    \label{fig:waterfall}
\end{figure*}
\subsubsection{2020 October 08 events}
 The events FRB~20201008B, FRB~20201008D and FRB~20201008C  were detected in the main lobe of \chimefrb~on 2020 October 8 within a time window of 3 seconds.  The first and second bursts were separated by 1.949 s while the time gap between the second and third burst was 0.954 s, indicating that all three bursts occurred within one 3.24-s \sgr1935 rotation period. The three detections, along with their preliminary estimates of $DM$ and fluence, were first reported in an Atel \citep{g++20}. Although the sub-leading bursts FRB~20201008D, and FRB~20201008C are orders of magnitude dimmer than the leading burst FRB~20201008B and typical FRBs in general, the fact that they are temporally very close to the bright FRB-like FRB~20201008B and happen within a rotational period of the magnetar, argues in favor of classifying them as FRB-like. We therefore use names assigned by the Transient Name Server\footnote{\url{https://www.wis-tns.org/}} (TNS).

When modeling these bursts, we chose to treat them as independent events and model them individually. We note however that regarding them as independent events as opposed to a single event with multiple components is somewhat ambiguous.  As was the case with \frb200428, FRB~20201008B is also an extremely bright event and our fiducial model is a poor fit to the data corresponding to the highest S/N detection beam. Therefore, we apply our model to a more amenable data from a different beam.  

For FRB~20201008C, a single-component fit to the data suggested a complex morphology, possibly indicating the presence of multiple sub-components. The $DM$ and width retrieved from the fit were considerably different from other \sgr1935 bursts. Consequently, we chose to apply a two-component model to FRB~20201008C. For this fitting, we employed a relatively strong yet a reasonable prior on $DM$ and width. Given its low S/N, we only fit up to the basic model with fixed $DI$ and $SI$. The results from this fitting are reported in Table \ref{tab:properties}. 

\subsubsection{Remaining two bursts from 2022} 
The remaining two events, FRB~20221014A and FRB~20221201A were observed on 2022 October 14 and 2022 December 01, respectively. These events were detected in the side-lobes, at hour angles of $-$99.8$\degree$ and $-$11.1$\degree$ respectively with estimated fluence of $\mathcal{O}$(10 kJy ms) as detailed later in \S\ref{sec:flux_fluence}. Given this high fluence, we register them as FRBs in the Transient Name Server. We model the highest S/N beam data of both these bursts under the fiducial model. The best-fit point estimate of parameters, based on the median of the converged MCMC chain, are reported in Table \ref{tab:properties} while the dynamic spectrum is included in Figure \ref{fig:waterfall}. %

\begin{figure*}[h!]
    \centering
    \includegraphics[scale=0.48]{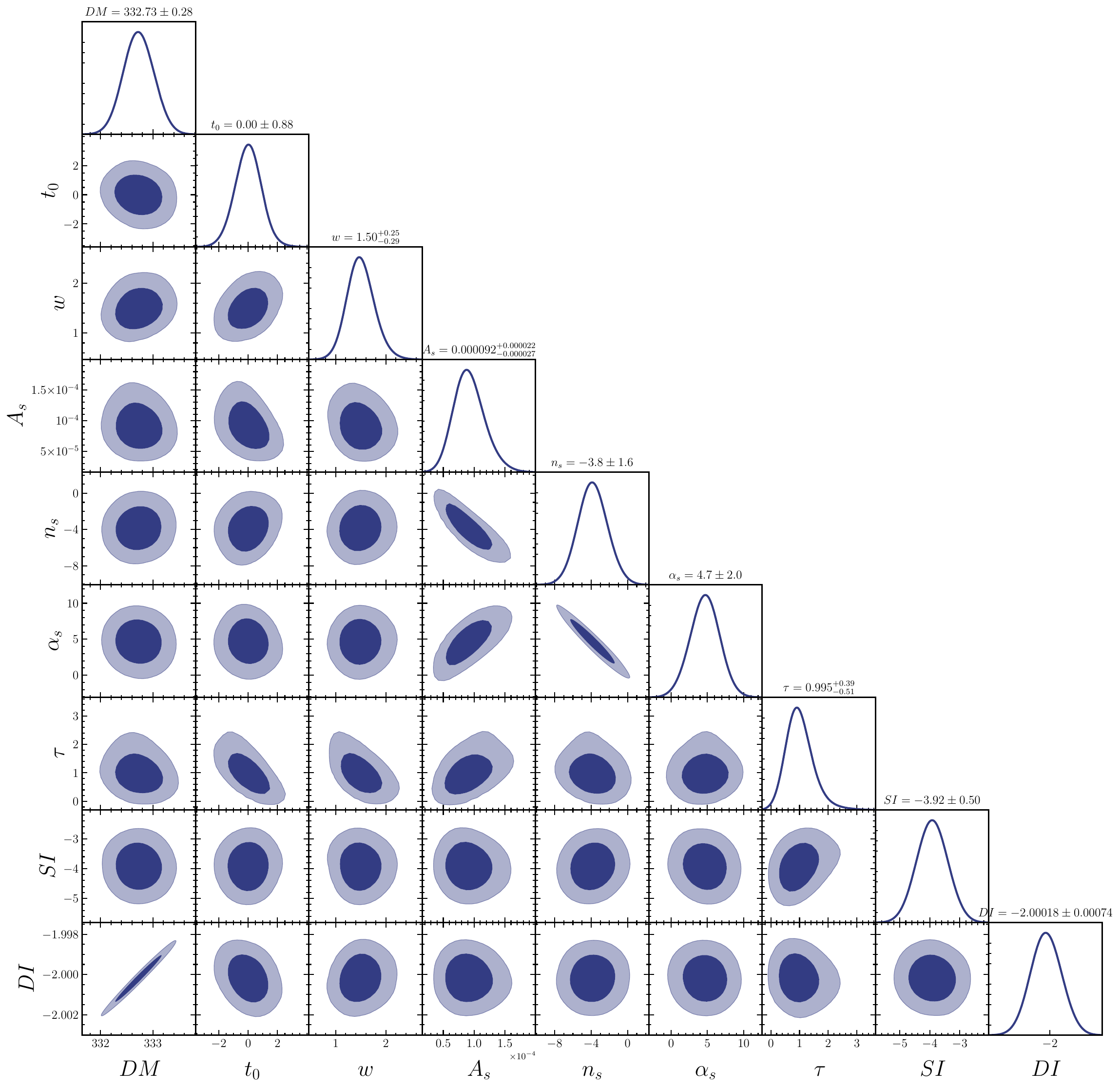}
    \caption{A triangle plot showing the covariance between parameters pairs as well as marginalized posterior distribution for all the parameters of our fiducial burst model for FRB~20221014A. The parameter $t_0$ is with respect to the arrival time of the burst at $\nu_{ref}=400.195$ MHz. All time measurements are in millisecond units. $DM$ is in $\mathrm{pc~cm^{-3}}$.}
    \label{getdist221014}
\end{figure*}

For the radio burst observed on 2020 October 14, FRB~20221014A, independent X-ray instruments, GECAM and HEBS, and Konus-\textit{Wind} \citep{2022ATel15686....1F, 2022ATel15682....1W} reported a short X-ray burst association. The GECAM observation consisted of a single pulse in the 20-100 keV band with a duration of about 250 ms. The Konus-\textit{Wind} emission was observed in two instrument's energy bands: G1(20-80 keV) and G2 (80-320 keV). The Green Bank telescope (GBT), which was actively monitoring \sgr1935 at that time, observed at least five bursts with significant signal-to-noise during a C-Band session \citep{2022ATel15697....1M}. All five radio bursts were detected within a time span of 1.5 seconds, well within one rotation of the magnetar, but over a range of phases. We find that our estimated arrival time of 2022-10-14T19:21:39.130 (UTC, topocentric) for FRB~20221014A is contemporaneous with the short X-ray burst reported by GECAM and HEBS \citep{2022ATel15682....1W} and Konus-\textit{Wind} \citep{2022ATel15686....1F} and remains consistent with two of the brightest bursts seen by GBT within their 0.1s uncertainty as reported by \cite{2022ATel15697....1M}. A higher precision estimate of the  X-ray arrival time can establish or rule out simultaneity of the two events. In order to aid such a comparative analysis in the future, we present parameter covariance result from our MCMC analysis of FRB~221014A in Figure \ref{getdist221014}. %

\subsection{Properties \& Morphology} 

The dispersion measures as well as scattering measures of the bursts analyzed in this paper and reported in Table \ref{tab:properties} are close to each other and do not show any secular evolution, suggesting no drastic evolution in the properties of the intervening medium over the 2.5-yr period. The dispersion indices of all the bursts, which are fit using a weak Gaussian prior, are within 2-3$\sigma$ of the theoretically expected value of $-2$. Except for FRB~20221201A, scattering indices are also found to be consistent with the theoretically motivated value of close to $\sim-4$, when fit with a Gaussian prior of width $\sigma=0.5$ centered around the value of $-4$.

None of the bursts show any hint of the `sad trombone' features often seen in repeating
FRB spectra \citep[e.g.][]{hss+19,abb+19c}. Moreover, all these events are quite broadband. This is of note as statistical studies of FRB population by \citet{Pleunis:2020vug} found repeater events to be on average broader in temporal width and more narrow-band than apparent non-repeaters.

\subsection{Localization}

The detected bursts were localized using the intensity localization pipeline of \chimefrb. This pipeline fits a model of \chimefrb's beams and an underlying spectral model to the spectra from all beams that detected the burst as well as all beams adjacent to those detection beams. For the events detected in CHIME's far side-lobes, FRB~20221014A and FRB~20221201A, we used a method identical to that presented by \citet{lin2023fast}. That is, we use a power-law model for the underlying burst spectrum and include only the FFT-formed synthesized beam model. For the events detected in CHIME's main lobe, FRB~20201008B, FRB~20201008D, and FRB~20201008C, a power-law model was also used, but the primary beam model was included. Table \ref{tab:burst_localization} shows the resulting burst localizations and their uncertainties. The positional uncertainties are derived from the statistical uncertainty summed in quadrature with the systematic errors estimated by \citet{lin2023fast}: 0.07 deg in RA and 0.10 deg in Dec. These burst positions are all consistent with the known position of \sgr1935 \citep{ier+16}.

\begin{table}[h!]
\begin{center}
\hspace*{-2.5cm}
\begin{tabular}{|c|c|c|}
	\hline
     TNS Name  & RA        &  Dec      \\
               &  deg       &  deg \\
     \hline\hline
  FRB~20221201A    & $293.75\pm0.06$   & $21.86\pm0.06$   \\
  FRB~20221014A    & $293.84\pm0.18$   & $21.84\pm0.17$   \\
  \multirow{2}{*}{FRB~20201008A}    & $293.8\pm0.2$   & $21.86\pm0.11$   \\
             & $292.0\pm0.1$   & $21.85\pm0.12$   \\
  FRB~20201008B    & $293.8\pm0.2$   & $21.88\pm0.11$      \\
  FRB~20201008C   & $293.75\pm0.06$   & $21.87\pm0.05$   \\
	\hline
\end{tabular}
\end{center}
\caption{The fitted sky positions for the five new \sgr1935 events.}%
\label{tab:burst_localization}
\end{table}

\subsection{Flux and Fluence estimation}
\label{sec:flux_fluence}
Lower limit fluence and flux values are estimated for each burst using the automated intensity flux calibration pipeline described by \cite{bridget_frb_fluence}. In brief, the spectrum of each burst is calibrated for flux and beam response using a steady source transit located closest in declination and time. In this automated analysis, we calculate the burst flux assuming that each burst was detected at ``beam boresight'', which we take to be the transit location of the burst position along the local meridian. Thus, these flux measurements are biased low, as bursts off-meridian will experience beam attenuation that is not accounted for. Note that, although the location of \sgr1935 is precisely-known, we do not use the method in \cite{aab+20} to scale the fluence by the beam response and obtain non-lower-limits for the bursts in this sample that were detected in the main lobe of the primary beam. Each of these bursts, although in the main lobe of the primary beam, are outside of the $600$\,MHz FWHM of the formed beam, meaning that parts of the bandwidth will be attenuated to the noise floor by the beam response (for a description of the formed beam response, see Section~2.2.1 of \citealt{bridget_frb_fluence}).

For FRB~20221014A and FRB~20221201A that were detected in the far side-lobe region \citep{abb+20, lin2023fast, lin2023constraints}, we complete additional analysis to obtain non-lower-limit estimates of the flux. 
\citet{lin2023constraints} mentioned an approach to calibrate the S/N of the side-lobe event. Here we use the similar approach to calibrate the flux,
\begin{equation}
    \mathrm{Flux} = \mathrm{\frac{G}{B}}\mathrm{Flux_{main}}, \label{eq:sidelobe_flux} \\
\end{equation}
where G is the geometrical factor in the range of 4-5, B is the beam response at the given hour-angle (HA) \citep{lin2023constraints}, and Flux$_\mathrm{main}$ is the flux reported by the automated pipeline assuming that the event was detected along the meridian in the main lobe. 
\begin{figure*}
    \centering
    \subfigure{\includegraphics[height=0.44\textwidth, width=0.48\textwidth, keepaspectratio]{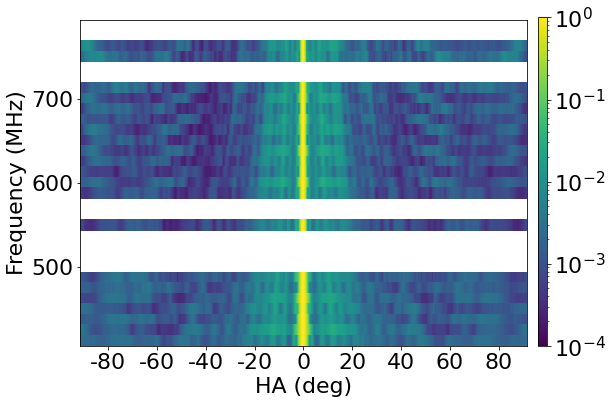}}
     \subfigure{\includegraphics[height=0.44\textwidth, width=0.48\textwidth, keepaspectratio]{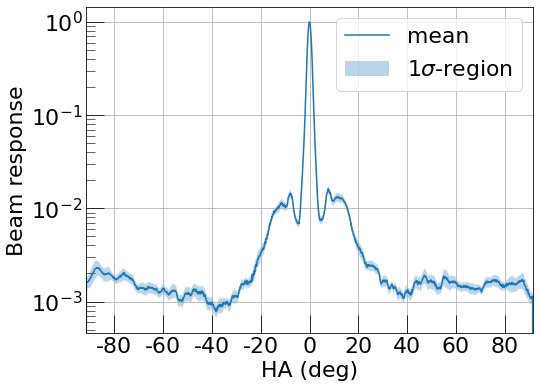}}
    \caption{The left panel shows the map of beam response at $400-725$ MHz obtained using the Crab Nebula as calibrator for HAs ranging from $-$91.5 to 91.5 deg. Channels contaminated with RFI have been masked manually. The right panel shows the frequency-averaged beam-response of the left panel.}
    \label{fig:beam_response}
\end{figure*}
We measure the beam response B using holographic measurements of the Crab Nebula by following the pre-processing procedures mentioned in \citet{lin2023constraints}, in which Figure \ref{fig:beam_response} shows the 2D and 1D beam response with the HA from $-$91.5 to 91.5 deg. To estimate the uncertainty of the beam response, we use the standard deviation of the frequency-averaged beam-response values 10 deg within the given HA. For FRB~20221201A with HA of $-$11.1 deg, the beam-response is 0.011$\pm$0.006. For FRB~20221014A, the HA of $-$99.8 deg is outside of the HA of holographic data (up to $-$91.5 deg). We assume the beam response is constant outside of the boundary, and the beam-response at $-$99.8 deg is the same as the beam response of 0.0015$\pm$0.0001 at $-$91.5 deg. The result is consistent with the solar beam measurement \citep{2022ApJ...932..100A}. We summarize the beam response measurements in Table \ref{tab:burst_ha_sidelobe}.

The flux and fluence of the two side-lobe events are on the order of $\sim$1 kJy and $\sim$10 kJy ms, respectively, and we summarize the results in Table \ref{tab:properties}.

\begin{table}[h!]
\begin{center}
\hspace*{-2.5cm}
\begin{tabular}{|c|c|c|}
	\hline
     TNS Name  & HA        &  B      \\
               &  deg       &  \\
     \hline\hline
  FRB~20221201A  & $-$11.1 &  0.011$\pm$0.006   \\
  FRB~20221014A  & $-$99.8 &  0.0015$\pm$0.0001    \\
	\hline
\end{tabular}
\end{center}
\caption{The hour-angle and the beam-response for the far side-lobe events.}
\label{tab:burst_ha_sidelobe}
\end{table}

\begin{deluxetable*}{c c c c c c c c}
\tabletypesize{\scriptsize}
\tablecaption{MCMC based estimates of model parameters for the radio bursts from \sgr1935. The estimates correspond to the median of converged MCMC samples. The uncertainty on time, flux, fluence and $DI$ are $1\sigma$ standard deviation of the samples. For DM, width, scattering and $SI$ we report the 68\% highest density interval.\label{tab:properties}}
\startdata
\\
    Properties
    & \multicolumn{2}{c}{FRB~20200428D \tablenotemark{a}}
    & FRB~20201008B 
    & FRB~20201008D 
    & FRB~20201008C \tablenotemark{b}
    & FRB~20221014A 
    & FRB~20221201A 
    \\\hline
    UTC Date & \multicolumn{2}{c}{2020-04-28} & 2020-10-08 & 2020-10-08 & 2020-10-08 & 2022-10-14 & 2022-12-01\\[1.0ex]
    Time \tablenotemark{c} & \multicolumn{2}{c}{14:34:24.4080(6)} {14:34:24.4370(5)} & 02:23:33.3578(8) & 02:23:35.307(1) & 02:23:36.261(3) & 19:21:39.130(2) & 22:06:59.0762(6) \\[1.0ex]
    DM [pc~cm$^{-3}$] \tablenotemark{d}& \multicolumn{2}{c}{$332.77^{+0.08}_{-0.08}$} & $332.50^{+0.13}_{-0.11}$ & $332.09^{+0.19}_{-0.20}$ & $332.67^{+0.12}_{-0.13}$ & $332.72^{+0.26}_{-0.30}$ & $333.11^{+0.09}_{-0.08}$ \\[1.0ex]
    Width [ms] & \multicolumn{2}{c}{$0.66^{+0.05}_{-0.05}$} {$0.33^{+0.07}_{-0.08}$} & $0.23^{+0.02}_{-0.04}$ & $0.29^{+0.08}_{-0.11}$ & $1.03^{+0.92}_{-0.97}$ & $1.48^{-0.25}_{-0.25}$ & $2.02^{+0.05}_{-0.04}$\\[1.0ex]
    Peak Flux  & \multicolumn{2}{c}{110 kJy}{150 kJy} & 266 $\pm$ 64 {Jy} & 16.3 $\pm$ 3.9 {Jy} & 1.8 $\pm$ 0.7 {Jy} & 2.1 $\pm$ 1.4\;{kJy} & 3.8 $\pm$ 3.0\;{kJy} \\[1.0ex]
    Fluence  & \multicolumn{2}{c}{420 kJy ms}{220 kJy ms} & 966 $\pm$ 239 {Jy ms} & 34.1 $\pm$ 8.1 {Jy ms} &  5.9 $\pm$ 1.7 {Jy ms}  &   9.7 $\pm$ 6.7\;{kJy ms} & 23.7 $\pm$ 18.0\;{kJy ms} \\[1.0ex]
    Hour Angle [deg] &  \multicolumn{2}{c}{22\degree} & -0.5\degree & -0.5\degree & -0.5\degree & -99.8\degree & -11.1\degree \\[1.0ex]
    Beam ID \tablenotemark{e} &  \multicolumn{2}{c}{2068} & 0062 & 0059 & 0059 & 0181 & 3061 \\[1.0ex]
    Scattering [ms]  \tablenotemark{f}& \multicolumn{2}{c}{$0.76^{+0.09}_{-0.10}$}& $0.72^{+0.08}_{-0.09}$ & $0.52^{+0.08}_{-0.09}$ & $1.13^{+0.5}_{-0.8}$ & $0.93^{+0.35}_{-0.45}$ & $0.92^{+0.11}_{-0.12}$\\[1.0ex]
    Dispersion index & \multicolumn{2}{c}{$-1.9998(2)$}&$-2.0004(3)$&$-2.0014(5)$&$-2$&$ -2.000(1)$ &$-1.99934(23)$\\[1.0ex]
    Scattering Index & \multicolumn{2}{c}{$-3.96^{+0.35}_{-0.37}$} & {$-3.70^{+0.46}_{-0.34}$}& $-4.27^{+0.45}_{-0.45}$& $-4$ & $-3.92^{+0.45}_{-0.54}$& $-2.62^{+0.32}_{0.30}$ \\[1.0ex]
    \hline
\enddata
\tablenotetext{a}{Results are a for 2-component fit to data. For flux and fluence, we simply quote the estimates from \cite{abb+20} which uses a different masking and calibration approach.}
\tablenotetext{b}{We fit a 2-component basic model to the burst with DI and SI fixed and only report the values corresponding to the primary peak. }
\tablenotetext{c}{Arrival time is in UTC (topocentric) at infinite reference frequency.}
\tablenotetext{d}{The DM is derived from the dispersion slope using a dispersion constant of $K=241$ GHz$^{-2}$ cm$^{-3}$ pc s$^{-1}$ \citep{manchester_and_taylor,Kulkarni:2020wss}}
\tablenotetext{e}{\chimefrb~beam ID of the data analyzed}
\tablenotetext{f}{The scattering is with respect to pivot frequency of 600 MHz}

\end{deluxetable*}

\subsection{Radio Upper limits on X-ray Bursts}
\label{subsec:radio_upper_limit}
In addition to the radio bursts discussed above, there were over 156 more X-ray or $\gamma$-ray bursts from \sgr1935 detected by either \textit{Swift}/BAT, {\it XMM-Newton}, {\it NuStar}, Konus-\textit{Wind}, \textit{Fermi}/GBM, {\it Agile}, or {\it INTEGRAL} IBIS/ISGRI between 2022 May 25 and 2022 December 13 that were published in either public ATels or GCNs.\footnote{We note that some of these 156 bursts may be repeats e.g., a burst detected both by \textit{Swift}/BAT and \textit{Fermi}/GBM will be counted twice.}\footnote{This time range was chosen as this was the period in which the authors maintained notices of all X-ray and $\gamma$-ray bursts from SGR1935+2154.} Out of these 156 bursts, seven were within the field of view of \chimefrb~at the time of their high-energy emission, where we define the field of view as within $\sim$ 17 degrees from the meridian\footnote{This is the region in which the CHIME/FRB beam is well modeled and understood.}. However, of these seven high-energy bursts, three occurred during times of non-nominal \chimefrb~sensitivity, and hence were ignored for the analysis that follows. 

For the four bursts which occurred within the FOV of \chimefrb~and occurred at times of nominal \chimefrb~sensitivity \citep{2022GCN.32770....1F, 2022ATel15752....1P, 2022GCN.32832....1V,2022GCN.32764....1V}, we calculate an upper limit on FRB-like radio emission associated with the burst. To calculate an upper limit on FRB-like radio emission, we follow the methods presented in \citet{2023ApJ...954..154C}. In summary, we use an FRB nearby in declination to act as our S/N-to-flux calibrator. We use models of both the primary and formed beams at \chimefrb\ to account for any subtle positional differences between the FRB calibrator and \sgr1935, and use system sensitivity metrics to account for temporal system sensitivity differences between the time of the FRB and that of the high-energy burst from \sgr1935. We then account for differences in the sky temperature (and hence sensitivity) between the FRB location and that of \sgr1935 using the 2014 Haslam all-sky continuum map at 408 MHz \citep{haslammap}. Finally, we scale our results to a detection S/N of 10, an intrinsic radio burst width\footnote{This only affects our fluence radio limits.} of 10 ms, and an estimated scattering time of 0.41 ms (estimated using NE2001). We also account for the different time of arrivals between the X-ray burst and the radio burst assuming a DM of 332.699 pc cm$^{-3}$. 
The resulting upper limits
for the four bursts in question are provided in Table \ref{ta:upper limits}.

The three bursts from \textit{Swift}/BAT and \textit{Fermi}/GBM do not have readily available X-ray fluxes and fluences. However, for the X-ray burst detected on 2022 October 13 at 02:02:46, Konus-\textit{Wind} reported a preliminary burst fluence of 3.30 $\pm$ 0.11 $\times 10^{-6}$ erg/cm$^2$ and a burst flux of 16.9 $\pm$ 1.4 $\times 10^{-6}$ erg/cm$^2$/s in the 20-500 keV band. Thus, the upper limits on the radio-to-high-energy flux and radio-to-high-energy fluence ratios are $8.5 \times 10^{-9}$ and $5.5 \times 10^{-11}$, respectively, assuming a 400-MHz bandwidth for \chimefrb. This radio-to-fluence ratio upper limit is five orders of magnitude lower than the radio-to-fluence ratio of $4^{+4}_{-2} \times 10^{-6}$ between the radio detection and Konus-\textit{Wind} detection of \sgr1935 from 2020 April. 

While the burst on 2020 April 20 was particularly hard, with a peak energy in the Konus-\textit{Wind} band (20 - 500 keV) of 82$^{+12}_{-9}$ keV, the burst on 2022 October 13 had a lower peak energy of 33$^{+3}_{-2}$ keV. The peak energy of the burst on 2022 October 13 was similar to the accompanying high-energy counterpart detected by Konus-\textit{Wind} for the radio burst detected by \chimefrb\ and presented above on 2022 Oct 14. The peak energy of this high-energy counterpart was 40$^{+6}_{-6}$ keV, in agreement within 1$\sigma$ with the peak energy report for the burst on 2022 October 13. Thus, not all radio bursts from \sgr1935 require a particularly high peak energy for an accompanying X-ray burst, yet bursts of similar peak X-ray energy are not always accompanied by observable radio bursts. 

\begin{deluxetable*}{c c c c c c}
\tablecaption{Upper Limits on FRB-Like Radio Emission from \sgr1935 at the times of X-ray bursts occuring near CHIME's meridian \label{ta:upper limits}}
\startdata
\\
    Time\tablenotemark{a}
    & Instrument \tablenotemark{b}
    & Hour Angle \tablenotemark{c}
    & Flux \tablenotemark{d}
    & Fluence \tablenotemark{e}
    & $\eta$ \tablenotemark{f} \\
    (UTC) & & (degrees) & Jy & Jy ms & (unitless) \\
    \\\hline
    10-13-2022 02:02:46 & Konus-\textit{Wind} & $-$1.0 & $<$36 & $<$234 & $<$$5.5 \times 10^{-11}$\\
    10-13-2022 02:36:00 & \textit{Swift}/BAT & 7.2 & $<$2300 & $<$14000 & N/A\\
    10-14-2022 02:20:34& \textit{Fermi}/GBM & 4.4 & $<$4000 & $<$25000 & N/A\\
    10-15-2022 02:13:55& \textit{Fermi}/GBM & 3.7 & $<$1800 & $<$11000 & N/A\\
    \hline
\enddata
\tablenotetext{a}{Detection time at the respective high-energy instrument.}
\tablenotetext{b}{Instrument for the high-energy detection.}
\tablenotetext{c}{Hour angle at \chimefrb\ accounting for the DM delay at 400 MHz associated with a DM of 332.699 pc cm$^{-3}$.}
\tablenotetext{d}{Upper limit on the FRB-like radio flux in the 400-to-800 MHz band.}
\tablenotetext{e}{Upper limit on the FRB-like radio fluence in the 400-800 MHz band scaling the burst to a 10-ms width.}
\tablenotetext{f}{Unitless ratio between the radio and high-energy fluences assuming a 400 MHz bandwidth for \chimefrb.}

\end{deluxetable*}

\subsection{Exposure and Sensitivity}
\label{subsec:exposure_sensitivit}

Among the six CHIME/FRB detections, three bursts occurred during the transit of the source in the main lobe of the telescope. The exposure in the main lobe was calculated following the methodology outlined by \cite{aab+21}. This corresponds to a cumulative duration of 111.14 hours above a fluence threshold limit of 10.3 Jy ms, with a confidence level of 95\%, from 2018 August 28 to 2022 December 1.

To determine the fluence threshold, we employed FRB~20190107A\footnote{\url{https://www.chime-frb.ca/catalog/FRB20190107A}} from \cite{aab+21} as it has a similar declination to that of \sgr1935. Using the CHIME/FRB beam model\footnote{\url{https://chime-frb-open-data.github.io/beam-model/}}, we assessed the sensitivity along the transit path for both the reference FRB and \sgr1935. The sensitivity ratio between \sgr1935 and the reference FRB was 0.8. Subsequently, we scaled the fluence threshold of the reference FRB by this ratio to obtain the final value for \sgr1935.

For events detected in a sidelobe, we present an upper limit on the exposures, taking into account the operational uptime of the \chimefrb\ system. The system remained operational for a duration of 1191.45 days, starting from 2018 August 28, until the cutoff date of 2022 December 1, for this publication. On average, only 95\% of the synthesized beams were online, corresponding to 984 beams out of 1024 beams \citep{abb+18}. Consequently, the total operational duration is reduced to 1145 days, denoted as ``beam days'' in this analysis. The exposure in days from the source can be estimated as follows - 
\begin{equation}
    Exposure\,(in\,days) = \frac{\Delta H}{\cos \delta \times 360^{\circ}} \times beam \; days 
\end{equation}
Here, the first term represents the fraction of a day during which a source at a given declination $\delta$ is observable by CHIME within a specific hour angle range $\Delta H$. For \sgr1935, the furthest detection has an hour angle of $-$100.7 deg. As the beam response cannot be accurately characterized at this position, we consider the second furthest detection from zenith, FRB~20221014A, corresponding to an hour angle of 91.5 degrees. Thus, the total hour angle transit range amounts to 183 degrees. By substituting these values, we obtain an upper limit on exposure of 627 days at a fluence threshold of 10.2 kJy ms. More details on the analysis are presented by \cite{lin2023fast}. 

The fluence threshold in the side-lobe is three orders of magnitude higher than that in the main lobe. This disparity arises because side-lobes are solely sensitive to exceedingly bright detections. To determine the sensitivity threshold for burst FRB~20221014A, we utilized the following relation:
\begin{equation}
    Fluence~threshold\,(\mathrm{Jy~ms}) = f \times \frac{G}{B_R} \times \frac{8}{S/N}.
\end{equation}
Here $f$ is the lower limit on the main lobe fluence of the burst estimated using the methodology described by \citet{bridget_frb_fluence}, $G$ is the geometrical factor derived by \cite{lin2023constraints}, $B_R$ is the beam response at hour angle 91.5 degrees estimated using Tau A holography, details of which are discussed by \cite{lin2023constraints}, $S/N$ is the 
real-time ({\tt bonsai}) signal-to-noise ratio for this event \citep{aab+21} and 8 is the cutoff S/N for which we store total intensity data. %

We detected a total of three bursts in the side-lobes during the interval from 28 August 2018 to 1 December 2022, so the rate of FRB-like bursts from the source for fluences $\gtrsim$ 10 kJy is $0.005^{+0.082}_{-0.004}$ burst per day, where the uncertainty is the 95\% confidence interval assuming a Poissonian process. This suggests that very bright bursts from \sgr1935 are rare occurrences, even during times of great X-ray activity.  

\section{Discussion} 

 In this paper we report on additional radio bursts, following those found in 2020 April \citep{abb+20}, detected by \chimefrb\ from the Galactic magnetar \sgr1935. In total, we present six bursts observed on four different occasions between 2020 April and 2023 December, with at least one burst with fluence  $\gtrsim \mathrm{1 ~kJy~ms}$ on each of these occasions. The peak fluxes of these bursts span a remarkable range of five orders of magnitude.  Although the brightest bursts seen thus far from  \sgr1935 have luminosities in the range of those the least luminous FRBs known, as discussed by \cite{aab+20},  FRBs as a class have luminosities over six orders of magnitude yet higher.  Thus it is yet unclear whether most FRBs are magnetars.  Still, the large dynamic flux range reported on in this work makes clear even a single magnetar is capable of a wide variety of radio burst luminosities.

We present a novel MCMC burst fitting code that has been developed for current and future precision FRB science needs, and use it to confirm that even when allowing both $DI$ and $SI$ to vary, the 2020 April FRB-like radio burst from \sgr1935 had both radio sub-bursts leading their X-ray counterparts by a few ms, far less than the rotation period of the magnetar, ruling out models that predict X-ray emission to precede or be coincident with the radio emission of this event.  For example, in a model that invokes synchrotron maser emission from decelerating relativistic blast waves to produce FRBs, \citet{metzger2019fast} and \citet{Margalit:2020luq} argue that X-ray emission arises from incoherent synchrotron emission from the same location as the radio burst \citep[i.e. in the downstream shock; see also][]{lyu14,bel20}.  They therefore predict near-coincident X-ray and radio bursts, or at most under a radio burst width apart.  However the X-rays in this model are optically thin at the shock deceleration radius, in contrast to the radio photons which are optically thick to induced scattering at the same location, hence escape after the X-rays. This is hard to reconcile with the results we have obtained in our analysis.  On the other hand, \citet{lp20} argue that FRBs are reconnection events in magnetar magnetospheres \citep[see][]{lyu02,wt19,lyu20} and predict that any radio burst must lead an associated X-ray burst by at most a few milliseconds, because it is only early in the reconnection event that the radio waves can escape.  This prediction is consistent with our results.  However,
 \citet{bel21,bel23} argue that such luminous radio bursts are damped within the magnetosphere and cannot escape; along the magnetic axis, the radio waves interact with plasma particles and accelerate them to high Lorenz factor at the expense of the wave energy.
 If so, FRB emission cannot originate therein.  Meanwhile, \citet{qkz22} argue that in the open field line region of magnetar magnetosphere, both the likely high outward speed of the plasma together with alignment of the field with the radio propagation direction reduce the expected interaction between the radio waves and the plasma, mitigating damping.  If so, then the close precedence of the radio to the X-ray bursts may indeed indicate a magnetospheric origin.

From a morphological point of view, the reported bursts from \sgr1935 appear different from typical repeating FRBs \citep[e.g.,][]{hss+19,abb+19c}, which tend to have broader bandwidths than the average for the repeaters \citep{pgk+21}.  Also, we find no evidence for downward frequency drifting (the ``sad trombone'' effect) so common to repeating FRBs.  However, small number statistics might be at play, given that bursts from repeaters can be broadband and exhibit no drifting. A clear prediction of the implication that at least some FRBs are magnetars \citep{aab+20,Bochenek:2020zxn}
is that narrow-band and/or frequency drifting will one day be seen in radio bursts from \sgr1935 and/or other established magnetars. \commentout{\utkarsh{Can't it be that FRBs have multiple sources including magnetar and the emission from magnetar does not show this feature? In that case we will never see downward drifting}}

In this work we find that \chimefrb, by virtue of its wide side-lobes, has excellent daily exposure to \sgr1935 and other Galactic magnetars for bursts above a fluence threshold of $\mathrm{10~kJy}$ ms when above the horizon.  Our estimated radio burst rate for \sgr1935,  is $0.005^{+0.082}_{-0.004}$ bursts per day for fluences $\gtrsim$ 10 kJy ms in the CHIME 400-800-MHz band in the period 28 August 2018 to 1 December 2022, suggesting that only 1--2 such bright bursts on average will be observable by CHIME/FRB per magnetar each year, and this is likely a strong upper limit since this rate was measured during a period of intense activity by \sgr1935.  Indeed although there are 10 catalogued Galactic magnetars known in the declination range covered by CHIME/FRB \citep[see][]{ok14}\footnote{\url{https://www.physics.mcgill.ca/~pulsar/magnetar/main.html}}, radio bursts from only \sgr1935 have been detected thus far by CHIME/FRB.  This suggests that the per source long-term average bright radio burst rate per Galactic magnetar is far lower than the value reported above.  This information from CHIME/FRB will be helpful for informing on the expected event rates of Galactic magnetar radio bursts for other planned wide-field radio instruments such as
GReX \citep{csk+21} and BURSTT \citep{lll+22} whose aim is to detect rare, high fluence short-duration radio bursts like those from SGR 1935+2154.

\begin{acknowledgments}
We acknowledge that CHIME is located on the traditional, ancestral, and unceded territory of the Syilx/Okanagan people.

We thank the Dominion Radio Astrophysical Observatory, operated by the National Research Council Canada, for gracious hospitality and expertise.

CHIME is funded by a grant from the Canada Foundation for Innovation (CFI) 2012 Leading Edge Fund (Project 31170) and by contributions from the provinces of British Columbia, Qu\'{e}bec and Ontario. The CHIME/FRB Project is funded by a grant from the CFI 2015 Innovation Fund (Project 33213) and by contributions from the provinces of British Columbia and Qu\'{e}bec, and by the Dunlap Institute for Astronomy and Astrophysics at the University of Toronto. Additional support was provided by the Canadian Institute for Advanced Research (CIFAR), McGill University and the Trottier Space Institute via the Trottier Family Foundation, and the University of British Columbia. The CHIME/FRB baseband system was funded in part by a CFI John R. Evans Leaders Fund grant to IHS. The Dunlap Institute is funded through an endowment established by the David Dunlap family and the University of Toronto. Research at Perimeter Institute is supported by the Government of Canada through Industry Canada and by the Province of Ontario through the Ministry of Research \& Innovation. The National Radio Astronomy Observatory is a facility of the National Science Foundation (NSF) operated under cooperative agreement by Associated Universities, Inc. FRB research at UBC is supported by an NSERC Discovery Grant and by the Canadian Institute for Advanced Research.  

U.G. thanks Mingyu Ge and Anna Ridnaia for their assistance with interpreting their data and answering some basic questions about their analysis. U.G. also thanks Daniel Foreman-Mackey for his suggestions regarding sampling-algorithms and Mark Halpern and Antonio Herrera Martin for their useful feedback. Support for this research was provided by the University of Wisconsin - Madison Office of the Vice Chancellor for Research and Graduate Education with funding from the Wisconsin Alumni Research Foundation. B.\,C.\,A. is supported by an FRQNT Doctoral Research Award. A.P.C is a Vanier Canada Graduate Scholar. V.\,M.\,K. holds the Lorne Trottier Chair in Astrophysics \& Cosmology, a Distinguished James McGill Professorship, and receives support from an NSERC Discovery grant (RGPIN 228738-13), from an R. Howard Webster Foundation Fellowship from CIFAR, and from the FRQNT CRAQ.	K.W.M. holds the Adam J. Burgasser Chair in Astrophysics and is supported by NSF grants (2008031, 2018490). K.R.S. acknowledges support from Fonds de Recherche du Quebec -- Nature et Technologies~(FRQNT) Doctoral Research Award.  M.B. is a Mcwilliams fellow and an IAU Gruber fellow. F.A.D is supported by the UBC Four Year Fellowship. B.M.G. is supported by an NSERC Discovery Grant (RGPIN-2022-03163), and by the Canada Research Chairs (CRC) program. C. L. is supported by NASA through the NASA Hubble Fellowship grant HST-HF2-51536.001-A awarded by the Space Telescope Science Institute, which is operated by the Association of Universities for Research in Astronomy, Inc., under NASA contract NAS5-26555.  M.M is supported by NSF grant 2307109 and DOE grant DE-SC0017647. A.B.P. is a Banting Fellow, a McGill Space Institute~(MSI) Fellow, and a FRQNT postdoctoral fellow. A.P. is funded by the NSERC Canada Graduate Scholarshops -- Doctoral program. Z.P. is a Dunlap Fellow. K.S. is supported by the NSF Graduate Research Fellowship Program. D.C.S. is supported by an NSERC Discovery Grant (RGPIN-2021-03985) and by a Canadian Statistical Sciences Institute (CANSSI) Collaborative Research Team Grant. S.P.T. is a CIFAR Azrieli Global Scholar in the Gravity and Extreme Universe Program.

\end{acknowledgments}

\facilities{CHIME}

\software{\textsc{matplotlib} \citep{matplotlib}, \textsc{numpy} \citep{numpy}, \textsc{emcee} \citep{Foreman_Mackey_2019}, \textsc{JAX} \citep{jax2018github}, \textsc{NumPyro} \citep{phan2019composable}, \textsc{getdist} \citep{lewis2019getdist}}, \textsc{astropy} \citep{astropy3}

\bibliographystyle{aasjournal}

\bibliography{frbrefs,frbrefs_utkarsh}

\end{document}